\documentclass[twocolumn,trackchanges]{aastex631}
\usepackage{multirow}
\usepackage{amsmath}
\usepackage{bm}
\usepackage{enumerate}
\usepackage{booktabs}

\defcitealias{poggianti+2017b}{P17b}
\defcitealias{peluso+2022}{P22}

\shorttitle{Gas-phase metallicity of local AGN in the GASP and MaNGA surveys}
\shortauthors{Peluso et al.}
\graphicspath{{Figures/}}

\newcommand{\oiii}{[\ion{O}{3}]}
\newcommand{\sii}{[\ion{S}{2}]}
\newcommand{\siii}{[\ion{S}{3}]}
\newcommand{\nii}{[\ion{N}{2}]}
\newcommand{\oi}{[\ion{O}{1}]}
\newcommand{\oii}{[\ion{O}{2}]}
\newcommand{\neiii}{[\ion{Ne}{3}]}

\begin{document}

\title{Gas-phase metallicity of local AGN in the GASP and MaNGA surveys: the role of ram-pressure stripping
}

\author[0000-0001-5766-7154]{Giorgia Peluso}
\affiliation{INAF-Padova Astronomical Observatory, Vicolo dell’Osservatorio 5, I-35122 Padova, Italy}
\affiliation{Dipartimento di Fisica e Astronomia "G. Galilei", Università di Padova, Vicolo dell’Osservatorio 3, 35122 Padova, Italy
}

\author[0000-0002-3585-866X]{Mario Radovich}
\affiliation{INAF-Padova Astronomical Observatory, Vicolo dell’Osservatorio 5, I-35122 Padova, Italy}

\author[0000-0002-1688-482X]{Alessia Moretti}
\affiliation{INAF-Padova Astronomical Observatory, Vicolo dell’Osservatorio 5, I-35122 Padova, Italy}

\author[0000-0003-2589-762X]{Matilde Mingozzi}\affiliation{Space Telescope Science Institute, 3700 San Martin Drive, Baltimore, MD 21218, USA}

\author[0000-0003-0980-1499]{Benedetta Vulcani}
\affiliation{INAF-Padova Astronomical Observatory, Vicolo dell’Osservatorio 5, I-35122 Padova, Italy}

\author[0000-0001-8751-8360]{Bianca M. Poggianti}
\affiliation{INAF-Padova Astronomical Observatory, Vicolo dell’Osservatorio 5, I-35122 Padova, Italy}

\author[0000-0002-5655-6054]{Antonino Marasco}
\affiliation{INAF-Padova Astronomical Observatory, Vicolo dell’Osservatorio 5, I-35122 Padova, Italy}

\author[0000-0002-7296-9780]{Marco Gullieuszik}
\affiliation{INAF-Padova Astronomical Observatory, Vicolo dell’Osservatorio 5, I-35122 Padova, Italy}

\begin{abstract}
Growing evidence in support of a connection between Active Galactic Nuclei (AGN) activity and the Ram-Pressure Stripping (RPS) phenomenon has been found both observationally and theoretically in the past decades.  In this work, we further explore the impact of RPS on the AGN activity by estimating the gas-phase metallicity of nuclear regions and the mass-metallicity relation of galaxies at $z \leq$ 0.07 and with stellar masses $\log {\rm M}_* / {\rm M}_\odot \geq 9.0 $, either experiencing RPS or not.
To measure oxygen abundances, we exploit Integral Field Spectroscopy data from the GASP and MaNGA surveys, photoionization models generated with the code {\sc Cloudy} and the code {\sc Nebulabayes} to compare models and observations. In particular, we build {\sc Cloudy} models to reproduce line ratios induced by photoionization from stars, AGN, or a contribution of both.
We find that the distributions of metallicity and \oiii~$\lambda$5007 luminosity of galaxies undergoing RPS are similar to the ones of undisturbed galaxies.
Independently of the RPS, we do not find a correlation between stellar mass and AGN metallicity in the mass range $\log {\rm M}_* / {\rm M}_\odot \geq 10.4$, while for the star-forming galaxies we observe the well-known mass-metallicity relation (MZR) between $ 9.0 \leq \log \ {\rm M}_* /{\rm M}_\odot \leq 10.8$ with a scatter mainly driven by the star-formation rate (SFR) and a plateau around $\log {\rm M}_* / {\rm M}_\odot \sim 10.5$. The gas-phase metallicity in the nuclei of AGN hosts is enhanced with respect to those of SF galaxies by a factor of $\sim$ 0.05 dex regardless of the RPS. 

\keywords{Galaxy environments --- Active Galactic Nuclei --- Galaxy chemical evolution}

\end{abstract}

\section{Introduction} \label{sec:intro}

The chemical evolution of a galaxy is regulated by a plethora of processes, from stellar winds and supernovae explosions within the galaxy body \citep[e.g.,][for a review]{larson+1974,larson+1975,maiolino-mannucci2019} to the exchange of material with its environment \citep[e.g.,][]{ellison+2009, peng+2014}. 
The global gas-phase metallicity is well-known to be strongly correlated with the assembled stellar mass of a galaxy \citep[e.g.,][]{lequeux+1979} through the so-called mass-metallicity relation (MZR) which has been
shown to hold from low $z$ \citep[e.g.,][]{tremonti+2004,perez-montero+2013} to high $z$ \citep[up to $z \sim$ 6.5 based on recent JWST measurements,][]{shapley+2023,curti+2023a,curti+2023b}.
At a given stellar mass, \cite{mannucci+2010} found for the first time an anti-correlation between the star-formation rate (SFR) and the metallicity which is the so-called Fundamental MZR (FMZR), while \cite{peng+2014} find that satellite galaxies in denser environments, in terms of local density, are more metal-rich than galaxies in lower-density environments.


In addition to stellar evolution and environmental effects,  also the presence of a central Active Galactic Nucleus (AGN)  potentially can have an impact on the galaxy metallicity \citep[e.g.,][]{groves+2006} and a relation
between the Narrow-Line Region (NLR) metallicity  and the (host galaxy) stellar mass has been investigated to test this hypothesis \citep[e.g.,][]{coil+2015, thomas+2019, dors+2020a, perez-diaz+2021, armah+2023}. While some studies do not find a correlation between these two quantities \citep{dors+2020b,perez-diaz+2021}, others do find a  relation both in the local universe \citep[e.g.,][]{thomas+2019, armah+2023} and at higher redshifts \citep[e.g.,][at $z \sim$ 3]{matsuoka+2018}.

The effect of the AGN on the metal content of the galaxy's central regions is also  highly debated:  regardless of the stellar mass of the host galaxy,
some works find that the AGN leads to an enrichment of metals, with AGN host galaxies showing higher central metallicity than star-forming galaxies of similar mass \citep[e.g.,][]{coil+2015,thomas+2019,perez-diaz+2021}, while other works measure lower metallicity in AGN than in star-forming regions \citep[e.g.,][]{donascimento+2022,armah+2023}. 

The origin of the metal enrichment may be explained by dust destruction in the Broad Line Region (BLR) which releases metals into the interstellar medium (ISM) \citep[][]{maiolino-mannucci2019} or \emph{in-situ} top-heavy IMF star-formation in the accretion disk  around the supermassive black hole (SMBH) \citep[e.g.,][]{nayakshin+2005, wang+2011}.
In the latter scenario, the AGN would foster rapid star formation and quick enrichment of the ISM, which in fact has commonly been observed to be very metal-rich \citep{maiolino-mannucci2019}. AGN-driven outflows of high metallicity gas, observed to be expelled on kpc scales from the BLR \citep[e.g.,][]{dodorico+2004,arav+2007}, would then enrich also the NLR. Another contribution to the metal enrichment of the gas surrounding the BLR may also come from \emph{in-situ} star-formation inside the AGN-driven outflows, which has been recently detected by several works \citep[e.g.][]{maiolino+2017,gallagher+2019}.



On the other side, a possible way to explain the low AGN metallicities measured by some other works is that AGN-driven winds halt the production of metals by quenching star formation in the circumnuclear regions around the galaxy center \citep{choi+2022}.
In support of this hypothesis,
\cite{armah+2023} find that the AGN X-ray luminosity-NLR metallicity relation anti-correlates with the Eddington ratio, which indicates that the low-luminous AGN (and therefore likely with the weakest feedback)  are more actively undergoing ISM enrichment through star formation, as opposed to the most luminous X-ray AGN. Similarly, EAGLE simulations predict that the scatter from the MZR at ${\rm M}_* > 10^{10.4} {\rm M}_\odot$ depends on the mass of the central black hole, and in particular that black hole mass and gas-phase metallicity are anti-correlated \citep{vanloon+2021}. 
AGN feedback can also play a role by removing both gas and metals from the nucleus of the galaxy and dispersing this material to larger radii \citep{choi+2022}.

In this context, a dedicated study of the gas-phase metallicity in the nuclear regions of galaxies hosting an AGN (AGN metallicity, hereafter) and its scaling relation with the host galaxy stellar mass in different environments
is still missing. 
A possible link between the AGN metallicity and the environment may have roots in the fact that environmental processes such as the ram pressure stripping (RPS) phenomenon have been proven to quench rapidly star formation in galaxies falling into clusters \citep{boselli-gavazzi+2006}, and that AGN feedback may aid in the quenching of star formation together with ram pressure \citep{ricarte+2020}. 
The {\sc ROMULUS C} cosmological simulations of a high-resolution galaxy cluster by \cite{ricarte+2020} find that RPS triggers enhanced gas accretion onto the black hole, which then produces heating and outflows due to AGN feedback. Growing evidence has been found, both observationally \citep{poggianti+2017b,peluso+2022} and theoretically \citep{tonnesen+2009,akerman+2023}, in support of the hypothesis that RPS is able to trigger or enhance the AGN activity in cluster galaxies. 
Recent studies have clearly identified AGN-driven outflows \citep{radovich+2019} and  AGN feedback in action \citep{george+2019} in strongly stripped galaxies. 

Another debated topic regards the choice of the more suitable method to derive the NLR metallicity \citep[e.g.,][]{dors+2015}. 
Similarly to what is typically adopted for the {\sc H ii} regions, the direct $T_{\rm e}$-method \citep[e.g.,][]{dors+2020a}, the strong emission-line (SEL) calibrators \citep[e.g.,][]{carvalho+2020,storchi-bergmann+1998} and photoionization models \citep[e.g.,][]{thomas+2018a, thomas+2018b, perez-diaz+2021} have been exploited in the literature to measure the NLR metallicity.
However, although AGN have high ionization degree, their high \citep[e.g.,][]{groves+2006} metallicity leads to faint auroral lines (such as the \oiii~$\lambda$4363), hampering 
the use of the $T_{\rm e}$-method.

Many SEL calibrations for the NLR have been computed in the last decades in the literature,
 derived either
with photoionization models \citep[e.g.,][]{carvalho+2020,storchi-bergmann+1998} or with the direct method \citep{flury-Moran+2020, dors+2021}. However, calibrators obtained from the same set of measurements (either direct or indirect) in case of ionization from star formation and AGN are still not available in the literature. 

Some works \citep[e.g.,][]{thomas+2019,perez-diaz+2021} have developed
consistent AGN and stellar photoionization models
and used them to determine the metallicity 
in both star-forming (SF) and AGN-ionized regions. Observed and predicted line ratios are compared making use of the Bayesian inference with codes such as {\sc Nebulabayes} \citep{thomas+2018a} or \textsc{{\sc H ii}-Chi-Mistry} \citep{perez-montero+2014,perez-montero+2019}.

In this paper, we  adopt an approach similar to \cite{thomas+2018a} to compute for the first time the gas-phase metallicity of the central regions of a sample of active galaxies affected by RPS. 
The aim is 
to look for signs of metal enrichment or decrement in the NLR, with respect to the case of ionization from star formation.
To do so, we draw galaxies from the GASP \citep[Gas Stripping Phenomena in galaxies,][]{poggianti+2017a} survey and from the MaNGA \citep[Mapping Nearby Galaxies at Apache Point Observatory][]{bundy+2015} survey. GASP is an ESO Large Programme carried out with the spectrograph MUSE to study galaxies in the local universe affected by RP in clusters.
The use of Integral Field Spectroscopy (IFS) allows us to derive the global metallicity by exploiting different extraction apertures and to spatially separate regions photo-ionized by stars or by the AGN. 

The paper is divided into the following sections. Section \ref{sec:data} presents the data sample and Section \ref{sec:models} presents the photo-ionization models that are used to compute the metallicity using the SEL method by comparing observed and predicted line ratios with the code {\sc Nebulabayes}, as described in details in Section \ref{sec:method}. Our results are presented in Section \ref{sec:results}: we investigate the effect of the aperture on our AGN metallicity measurements, and study how AGN metallicity correlates with different galaxy properties (stellar mass, AGN luminosity) using both RP stripped galaxies and a control sample of field galaxies that are not disturbed by the environment.
In Section \ref{sec:discussion}, we sum up and discuss the
results.

We adopt a \cite{chabrier+2003} initial mass function in the mass range 0.1-100 M$_\odot$. We assume a standard $\Lambda$CDM cosmology with $\Omega_m$ =0.3, $\Omega_{\lambda}$ = 0.7 and $H_0$ = 70 km $\rm s^{-1}$ Mpc$^{-1}$.







\section{Datasets and Galaxy samples} \label{sec:data}
The goal of this Section is to build four different samples to study the gas-phase metallicity of galaxies in different physical conditions, exploiting ancillary data from the GASP and MaNGA surveys.
To do so, we use the samples already presented in \cite{peluso+2022} (P22 hereafter) and \cite{vulcani+2018}.
In particular, we build a RPS sample of galaxies either hosting an AGN (AGN-RPS)  or not  (SF-RPS) and a control sample of galaxies located in the field (AGN-field sample and SF-field sample, i.e. AGN-FS and SF-FS hereafter), thus undisturbed by the RPS. 
All the galaxies are late-type and have ongoing star-formation in the galactic disk.
 To classify the ionization mechanism acting on the gas, we make use of the so-called BPT diagnostics \citep{baldwin+1981, veilleux+1987,kauffmann+2003,kewley+2001,kewley+2006}. 
Specifically, we use the BPT diagram involving the line ratios \nii$\lambda$6584/H$\alpha$ over \oiii$\lambda$5007/H$\beta$ (e.g., \nii-BPT)\footnote{The only exception is the GASP galaxy JW100  for which we use the \sii-BPT, involving the \sii$\lambda\lambda$6716,6731/H$\alpha$ instead of \nii$\lambda$6584/H$\alpha$, because at the galaxy's redshift the \nii~ line is contaminated by a sky line}. In this case, the \cite{kewley+2001} relation based on photoionization models is used to delimit the region where Seyfert/LINER spaxels are located, and the empirical \cite{kauffmann+2003} relation to isolate star-forming spaxels. The region in between the two demarcation lines is populated by spaxels with line ratios usually classified as Composite (SF+AGN). We finally use the \cite{sharp+2010} relation to further distinguish Seyfert from LINER line ratios.

\subsection{GASP sample} \label{sec:gasp-sample}

The GASP survey is a program focused on the study of gas removal processes due to the interaction between the intra-cluster medium (ICM) and the ISM. 
The survey observed 
114 galaxies at $0.04 < z < 0.07$ located in clusters, groups and the field,
with the integral-field spectrograph MUSE, mounted at the Very Large Telescope (VLT), which has a field of view of  $1' \times 1'$ and covers a spectral range from 4800 to 9300 \AA~ (rest-frame) with a median resolution FWHM $\sim$ 2.6 \AA. GASP observations were taken in Wide-Field Mode with natural seeing (WFM- noAO) with an average seeing of $\sim 1 \arcsec$. More details on the sample selection and data analysis can be found in \cite{poggianti+2017a}.


Stellar masses range from $10^{9}$ to 3.2 $\times 10^{11} {\rm M}_\odot$ and have been computed with the code {\sc SINOPSIS} \citep{fritz+2017} assuming a \cite{chabrier+2003} IMF. In brief, the code SINOPSIS uses a stellar population synthesis technique that reproduces the observed optical spectra of galaxies performing a spectral fitting of the stellar content and extinction, to derive the spatially resolved properties of the stellar populations. As in \cite{vulcani+2018}, stellar masses are obtained by summing the stellar mass computed with {\sc SINOPSIS} inside each spaxel within the galaxy.
 The emission lines are fitted with the code {\sc KUBEVIZ} \citep{fossati+2016}, from the continuum subtracted and extinction-corrected MUSE spectrum, as described in \cite{poggianti+2017a}. 

 Integrated galaxy global properties such as inclination, position angle, and effective radius (${R_{\rm e}}$) are measured from  I-band MUSE photometry, as described in detail in \cite{franchetto+2020}. In particular, the effective radius $R_{\rm e}$ was computed from the luminosity growth curve $L(R)$ of the galaxies, obtained by trapezoidal integration of their surface brightness profiles.

 For our analysis, we select the MUSE spaxels in which the following emission lines have S/N $>$ 3:
H$\beta$, \oiii~ $\lambda$5007, H$\alpha$, \nii~ $\lambda$6584, \sii~ $\lambda$6716 and \sii~ $\lambda$6731.
We exclude the \oi~ $\lambda$6300 emission line because the flux is often faint (i.e., S/N $<$ 3 in most of the spaxels of the galaxies) and photoionization models struggle to predict the emission of this line in agreement with the observations \citep[see e.g.,][]{law+2021, dopita+2013, dopita+1997}. 
Another reason to exclude the \oi~ line is given by
the \oi~ excess \citep{poggianti+2019}, typically observed in regions located in the  tails of RP-stripped galaxies.
Though the mechanism driving the \oi~ excess is not yet fully understood, it should mainly affect the outer regions of the emitting clouds, where \oi~ is formed.

From the GASP-RPS sample in P22, we select the 11 RP-stripped galaxies with Seyfert or LINER-like nuclei (AGN-RPS) according to the spatially-resolved BPT diagnostics and 39 star-forming RP-stripped galaxies without AGN activity (SF-RPS). We excluded four SF galaxies (JO95, JO156, JO153 and JO149) from the RPS sample in P22 as their irregular I-band morphology prevented a good estimate of their structural parameters \citep{franchetto+2020}, which were necessary to extrapolate the nuclear metallicities as described in details in Section \ref{sec:results}.


Moreover, from the GASP control sample in \cite{vulcani+2018}, we consider 15 galaxies located in the field and undisturbed by RP (i.e., SF-FS).
We exclude only one galaxy (P19482) from the original sample in \cite{vulcani+2018} which was found to be located in a filament in a subsequent work \citep{vulcani+2021}. 
\\
\\

\subsection{MaNGA sample} \label{sec:manga-sample}

 Mapping Nearby Galaxies at Apache Point Observatory \citep[MaNGA,][]{bundy+2015} is an integral-field spectroscopic survey using the BOSS Spectrograph \citep{smee+2013} mounted at the 2.5 m Sloan Digital Sky Survey (SDSS) telescope \citep{gunn+2006},  covering a spectral range from 3600 to 10300 \AA~ at R $\sim$ 2000. Briefly (we refer the reader to P22 for more details) we exploit the MaNGA Data Release 15  \citep[DR15,][]{bundy+2015} and, in particular,
we use the Pipe3D-v2\_4\_3\footnote{\url{https://www.sdss.org/dr16/data_access/value-added-catalogs/?vac_id=manga-pipe3d-value-added-catalog:-spatially-resolved-and-integrated-properties-of-galaxies-for-dr15}}  \citep{sanchez+2016,sanchez+2016b, sanchez+2018} catalog to select star-forming (i.e., with specific star-formation rate, sSFR, $\rm > 10^{-11} yr^{-1}$) galaxies and the visual morphological classification from the MaNGA Value Added Catalogs\footnote{\url{https://www.sdss.org/dr16/data_access/value-added-catalogs/?vac_id=manga-visual-morphologies-from-sdss-and-desi-images}} \citep{hernandez+2010} to select late-type galaxies with 0 $<$ TType  $<$ 12. To ensure that  galaxies are not affected by RPS, we consider only galaxies located in halo masses lower than $M_{{\rm halo} < 10^{13} } \ {\rm M}_\odot$, according to the \cite{tempel+2014} environmental catalog, that will be referred to as filed galaxies. In this way, we obtain the MaNGA control sample of 782 galaxies presented in P22 with similar properties (such as morphology and sSFR) to the GASP RPS sample but most likely undisturbed by environmental processes.

The P22 sample covers a redshift range $0.0024 < z < 0.1439$, within which the MaNGA spatial resolution corresponds to a physical size that goes from 0.13 kpc to 6.31 kpc. In fact, the IFU fiber size is 2$\arcsec$ and the reconstructed PSF inside the IFU
has an FWHM of 2.5$\arcsec$, which corresponds to the spatial resolution of the observations \citep{law+2016}.  

To have approximately a similar spatial resolution in MaNGA and GASP (i.e., $\sim$ 1 kpc), 
we select the 530 MaNGA galaxies with redshift $z \leq 0.04$. 
In this way,  30\% of the galaxies are at $z \sim$ 0.025, where the spatial resolution corresponds to $\sim$ 1 kpc, reaching at most 1.98 kpc of resolution ($z=0.04$), which is still within a factor of two with respect to the resolution of GASP. 


Among the 530 galaxies, we include in our sample only those (444/530) belonging to the Primary+Color-Enhanced sample \citep{bundy+2015}, in order to have a smooth distribution in redshift, while ensuring complete coverage of the i-band magnitudes.
The Primary+Color-Enhanced sample covers uniformly the galaxy up to 1.5 times the effective radius (1.5 $R_{\rm e}$), which perfectly suits our purposes, as we aim at characterizing the galaxy's central regions.

Finally our sample is further reduced to the 429 galaxies for which we were able to extract the aperture-corrected stellar mass from the Principal Component Analysis (PCA) catalog \citep{pace+2019a,pace+2019b}. In particular, the aperture correction needed to take into account the galaxy mass residing in the region extending outwards with respect to the 1.5 $R_{\rm e}$ aperture is recovered 
with the Color-Mass-To-Light Relations method, which employs relations \citep[as the one in][] {pace+2019a} between the mass-to-light ratio and the photometric colors of the galaxy's light outside the MaNGA IFU. 

The final sample spans the stellar mass range 9.0 $\leq \ \log ({\rm M}_* /{\rm M}_\odot) \ \leq$ 11.3. Among the 429 galaxies, 52 are Seyfert/LINER (i.e., AGN-FS) according to the spatially-resolved BPT classification with the \nii /H$\alpha$ versus \oiii / H$\beta$  diagnostic \citep[NII-BPT,][]{baldwin+1981}. 377 galaxies are instead classified as star-forming and were added to the SF-FS. 
As galaxies part of the SF-FS are observed either by the GASP and MaNGA surveys, we are able to check the consistency of our estimates of metallicity or stellar masses ensuring that our measurements are not affected by systematic effects (e.g., different data reduction and instruments) caused by the use of two surveys.

We use the online tool MARVIN\footnote{\url{https://www.sdss.org/dr16/manga/marvin/}} \citep{cherinka+2019} to download both the de-projected coordinates of our targets (\emph{spx\_ellcoo}) and the emission line fluxes (\emph{gflux}). The de-projected coordinates are
computed using the ellipticity ($\epsilon$ = 1-b/a) and position angle ($\theta$) measured from the r-band surface brightness. 
The same emission line fluxes listed in Section \S \ref{sec:gasp-sample} are drawn from the drpall-v2\_4\_3 and have S/N $>$ 1.5, which is the value typically adopted in MaNGA  \citep{belfiore+2019}. 
The emission lines are fitted with a Gaussian function and are corrected for stellar absorption, since the Data Analysis Pipeline \citep[DAP;][]{westfall+2019,belfiore+2019}
simultaneously fits the continuum and emission lines with the latest version of the pPXF software package \citep{cappellari+2017}. 
All lines are also corrected for Galactic extinction, using the \cite{schlegel+1998} maps \citep{westfall+2019} and the reddening law of \cite{o'donnell+1994}.  Following the same approach used in GASP, we correct the emission lines for host galaxy dust attenuation using the \cite{cardelli+1989} law and assuming an intrinsic Balmer decrement $\rm I(H\alpha)/I(H\beta) = 2.86$, appropriate for an electron density $n_{\rm e}$=100~cm$^{-3}$ and electron temperature $T_{\rm e} =10^4~K$ \citep{Osterbrock2006}.

\section{Photoionization models} \label{sec:models}

Models are generated with {\sc Cloudy} v17.02 \citep{ferland+2017}
in case of ionization from stars ({\sc H ii} models, hereafter) and AGN (AGN models, hereafter), so that the metallicity is measured in a homogeneous way from the central AGN region to the star formation-dominated outskirts of galaxies with AGN activity. To compute the metallicity in Composite (AGN+SF) regions, we mix the {\sc H ii} and AGN models as described in detail in Section \S \ref{sec:method}.

The files used as input by  {\sc Cloudy}  are built using the {\sc CloudyFSPS} library \citep{byler+2018}, modified to handle both {\sc H ii} and AGN models.
All models span the following parameter space:

\begin{itemize}
\item The ionization parameter [$\log(U)]$ ranges between $ -4 \leq \log (U) \leq -1$ with a step of 0.5 dex;
\item Gas-phase abundances are those   in {\sc CloudyFSPS} that are based on the  solar values from \citet{dopita+2000} 
\citep[see][]{byler+2017}. 
With the exception of nitrogen and helium, abundances scale with the gas-phase metallicity ($\log Z$ = $-1, -0.6, -0.4, -0.3, -0.2,$ $ -0.1, 0.0, +0.1, +0.2, +0.3, +0.4, +0.5$),  corresponding to oxygen abundance ranging between $\rm 7.69 \leq 12 + log \rm (O/H) \leq 9.19$ (12 + log (O/H) = 8.69 for the solar value). For nitrogen and helium, the relations with $\log Z$ are those defined in \citet{dopita+2000} to take into account the effects of non primary nucleosynthesis. The effects of abundance depletion by grains is also taken into account in  {\sc CloudyFSPS}, following \citet{dopita+2000}.
\end{itemize}

We run our grids of {\sc Cloudy} models, iterating till the temperature is above 100 K or until convergence: since in the outer regions the ionization rate may fall below the galactic background rate, cosmic ray background emission \citep{ferland+1984} was added as a secondary ionization source.
We explored the effect of dust on the line ratios studied here by comparing models with and without dust grains. To this end we assumed for the grains the default size distribution and abundances in the diffuse interstellar medium of our galaxy \citep{vanhoof+2001,vanhoof+2004,ferland+2013}, described by
the {\sc grains ISM} command in {\sc Cloudy}. 
Consistently with \citet{byler+2017} we find that its effect is minimal, with dusty models producing slightly higher \oiii~/H$\beta$ (i.e. $\sim$ 0.19 dex, on average) at fixed \nii~/H$\alpha$ for high metallicities and ionization parameters. 

We also explored the effect of varying the dust-to-metal abundance, without observing any significant effect. 

\begin{figure*}
     \makebox[\textwidth]{
     \includegraphics[scale=0.9]{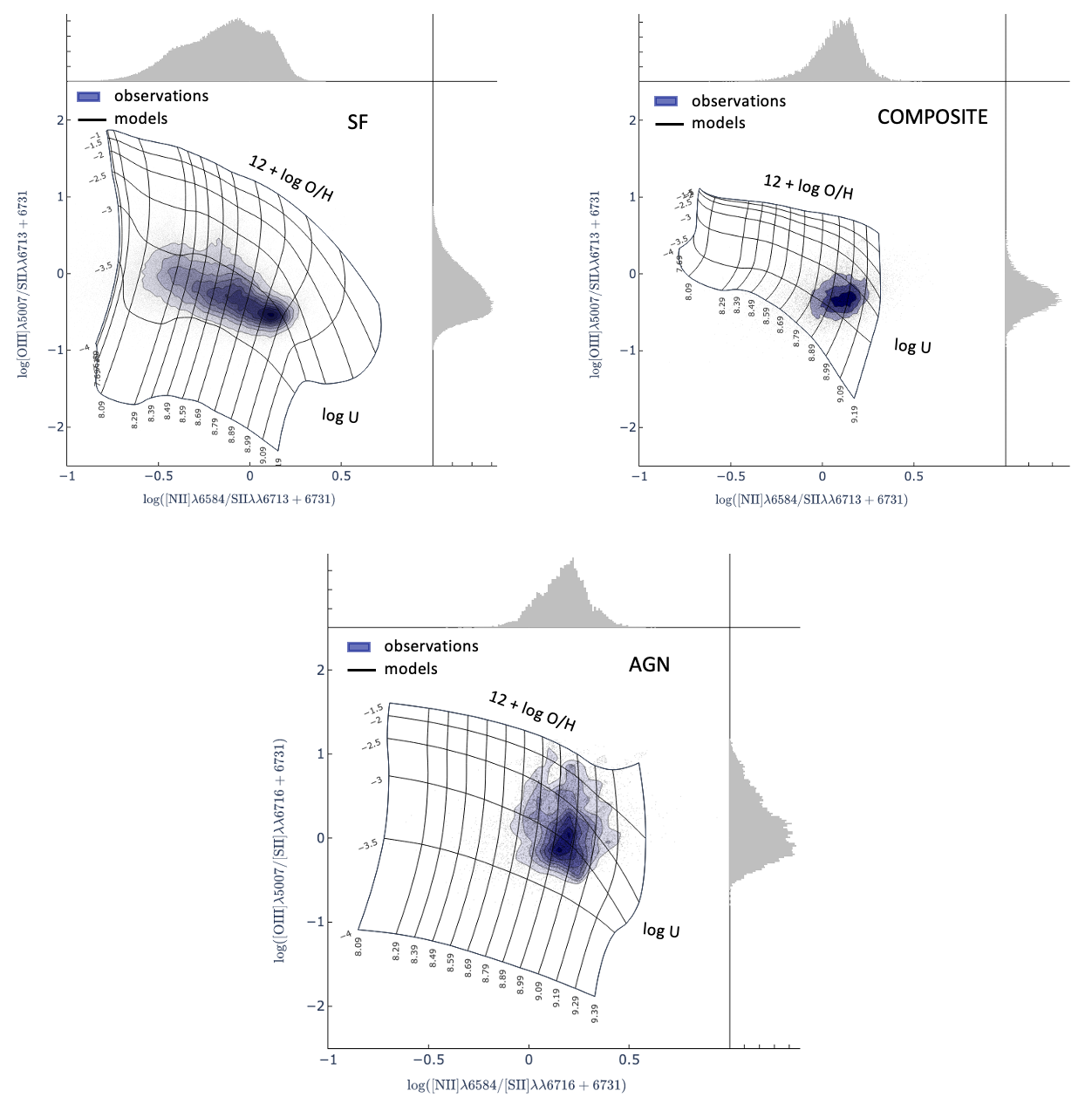}}%
      \caption{\oiii~/\sii~ vs \nii~/\sii~ line ratios in case of ionization from SF (top left), AGN+SF (top right), and AGN (bottom). 
      The grey points are the observed line ratios inside the spaxels of the MaNGA and GASP samples together. 
      The distribution of the observed points is  outlined by density curves filled with different shades of purple and shown by the grey histograms in the top and right insets. Darker colors indicate regions where the density of data is higher. The black solid lines are the {\sc Cloudy} models. 
      The {\sc H ii} models have stellar ages $t_*$ = 4 Myr, Composite models have the ionization parameter of the stars $\log U_{\rm {\sc H ii}}$ is fixed to -3.0 and $f_{AGN}$ is 0.2, AGN models have  $\alpha$ = -2.0 (see text for details).}
      \label{fig:models}
\end{figure*}

Finally, we select models with a gas density of ${\rm n}_H = 10^2 \ {\rm cm}^{-3}$ since they fully recover the observed line ratios in our GASP and MaNGA samples, as shown in Figure \ref{fig:models}. 
In particular, in Figure \ref{fig:models} we show the  line ratios \oiii~/\sii~ and \nii~/\sii~ of the SF (top left), Composite (top right), AGN (bottom) models and the observed line ratios inside the spaxels in MaNGA and GASP classified by the BPT correspondingly. 
The purple-shaded curves outline the density distribution of the observations, which is fully covered by the model grid.


\subsection{{\sc H ii} models}

{\sc H ii} models are generated following the same prescription as in \cite{byler+2017}. The python library {\sc python-fsps} is used to generate the ionizing continuum produced by a Single Stellar Population (SSP). To this end, we use the SSPs produced by the Flexible Stellar Population Synthesis code \citep[FSPS,][]{conroy+2009} 
and  the MESA Isochrones and Stellar Tracks \citep[MIST;][]{choi+2016,dotter+2016}. 
In {\sc Cloudy}, the stellar continuum models produced by {\sc FSPS} are read by the {\sc Table STAR} command, that also takes as input  the stellar age and metallicity\footnote{Stellar metallicities in MIST are defined within $-2.5 \le \log(Z/Z_\odot) < +0.5$.} to be used. For each {\sc Cloudy} model, the gas phase metallicity equals to the stellar metallicity.

Unlike the \cite{byler+2017} models, we use the version {\sc Cloudy} v17.02 due to several improvements in the atomic database introduced with respect to {\sc Cloudy} v13 \citep{ferland+2013}, in particular concerning the  rate coefficient for the S$^{2+}$ - S$^+$ dielectronic recombination \citep{ferland+2017, badnell+2015, belfiore+2022}. 


We test models with stellar ages ranging between 1 Myr $\leq t_* \leq$ 7 Myr \citep[similary to][]{byler+2017} , and we 
fix $t_* =$ 4 Myr \citep[see also][]{mingozzi+2020}, as  models with stellar ages  $t_* \leq$ 4 Myr are perfectly capable to reproduce line ratios typically observed in {\sc H ii} regions \citep[in agreement with e.g.,][]{dopita+1997}, while models with $t_* \geq$ 4 Myr generate line ratios (such as \oiii~/H$\beta$ and \nii~/H$\alpha$) too weak to reproduce the entire range of the observed MaNGA and GASP line ratios of our sample. 

\subsection{AGN models}
For AGN photoionization models, we adopt as ionizing source a simple power law continuum (command {\sc table power law}  in {\sc Cloudy}):

\begin{equation}
    \rm
    S_\nu \propto 
    \begin{cases}
    \nu^{\alpha}\ \ h \nu_1< h\nu < h \nu_2\\
    \nu^{5/2} \ \ h\nu<h\nu_1 \\
    \nu^{-2} \ \ h\nu>h\nu_2\\
    \end{cases}
\end{equation}

where $h\nu_1=9.12 \times10^{-3}$ Ryd and $h\nu_2=3676$ Ryd define the spectral breaks at 10$\mu$m and 50 keV respectively. The slope of the continuum, from the infrared to X-ray wavelength ranges, is set equal to $\alpha$ = - 2.0, as in this way the models were able to perfectly reproduce the observations.
According to the literature, the NLR density is relatively high \citep[i.e., $n_{\rm e} \approx  300-500 \rm \ cm^{-3}$,][]{storchi-bergmann+1998, feltre+2016, armah+2023} with respect to the {\sc H ii} regions in the  galaxy disc \citep[10 $\rm cm^{-3}$, e.g.,][]{dopita+2013}. However we note that the regions classified as AGN by the BPT can extend well-beyond the sub-kpc scale of the NLR \cite[i.e., the so-called extended narrow line region, ENLR, see e.g.][]{congiu+2017,chen+2019} thus we consider $n_{\rm H} = 10^2 \rm \ cm^{-3}$ as an average value between the high-density (i.e., 500 $\rm cm^{-3}$) and the low-density (i.e., 10 $\rm cm^{-3}$) regime. 




\subsection{Composite models}

We combine the {\sc H ii} and the AGN models following a similar approach as in \cite{thomas+2018b}. 

The mixed emission is parametrized by f$_{AGN}$, defined as:

\begin{equation}
\nonumber
f_{AGN} = \frac{R_{AGN}}{R_{{\rm HII}}+R_{AGN}}
\end{equation}

where $R$ is the flux of the reference line (i.e. H$\beta$), thus $R_{AGN}$ is the H$\beta$ flux that arises from the AGN and $R_{{\rm H II}}$ is the the H$\beta$ flux that arises from the {\sc Hii} regions. In other words, $f_{AGN}$ is the fraction of the H$\beta$ flux from the AGN with respect to the total H$\beta$ flux (i.e. $R_{{\rm H II}}+R_{AGN}$) of a mixed spectrum, where the emission comes from both stars and AGN.

We obtain the Composite grids with the following steps:

\begin{itemize}
\item[1.] we mix the {\sc H ii} and AGN models with the same metallicity and gas density;
\item[2.] 
the mixed emission line ratios are computed as:

\begin{equation}
\nonumber
\hspace*{-0.5cm}
    \left(\frac{L}{R}\right)_{Comp} = \left(\frac{L}{R}\right)_{AGN} \times f_{AGN} + \left(\frac{L}{R}\right)_{\rm H II} \times (1 - f_{AGN})
\end{equation}

where  
$L$ is the flux of a generic line.

\end{itemize}

The $f_{AGN}$ is a parameter of the Composite models, which ranges between 0.2 (i.e., 80 \% of ionization due to the stars) to 1 (i.e., 100 \% of ionizing photons coming from the AGN), with a step of 0.2.

In the Composite models, $\log(U)$ indicates the ionization parameter of the AGN emission, while the ionization parameter of the stars $ \log U_{\rm H II}$ is fixed to -3.0 (i.e., median value observed in pure SF regions), similarly to \cite{thomas+2018b}.

\section{METHODS} \label{sec:method}

In this section, we show how we set our {\sc Nebulabayes} \citep{thomas+2018a} analysis and derive the gas-phase metallicity and the ionization parameter. 
In brief, the code takes as input a set of emission lines from photoionization models and a set of observed emission lines with their relative errors. The line fluxes are then divided for a reference line, specified by the user. By comparing observations and predictions using the Bayes theorem, the code finds the best model to fit the observable.
{\sc Nebulabayes} is provided with models, generated with the code {\sc Mappings 5.1} \citep{sutherland-dopita+2017}, for both the {\sc H ii} and AGN-ionized regions. However, in Appendix \ref{sec:appendix-a} we discuss in detail the reason which lead us to generate and use our own {\sc Cloudy} models (presented in Section \S \ref{sec:models}).

To obtain the metallicity and ionization parameter computed spaxel-by-spaxel, we use our SF/Composite/AGN models inside the spaxels within the galaxy classified by the BPT diagrams correspondingly and compare the predicted and observed emission lines \oiii~ and \nii~ normalized for the reference line \sii. 
By using
the plane \oiii~/\sii~  (sensitive to $\log U$) 
versus \nii~/\sii~  (sensitive to $\log Z$), 
we are able to distinguish very well models with different values of ionization parameter and metallicity, as shown in Figure \ref{fig:models}. In this Figure, we plot the {\sc H ii}, AGN, and Composite models on the plane \oiii~/\sii~ vs. \nii~/\sii~ demonstrating the ability of such lines to unfold the grids.

The \oiii~/\sii~ ratio is sensitive to the variation in $\log (U)$ because of the different ionization potentials (IP) needed to create the O++ and  S+ ions (35.12 eV and 10.36 eV, respectively). Instead, the ions emitting the \nii~ and \sii~ lines have similar IP and thus the ratio \nii~/\sii~ has little dependence on $\log(U)$. 
However, the \nii~/\sii~ is a  good indicator for  Z as the growth of N/H scales with Z$^2$ \citep{hamman-ferland-1993} while S/H is $\sim$ Z \citep[e.g.][]{dors+2023}.

\begin{figure}[ht]
    \makebox[0.4\textwidth]{
    \includegraphics[scale=0.8]{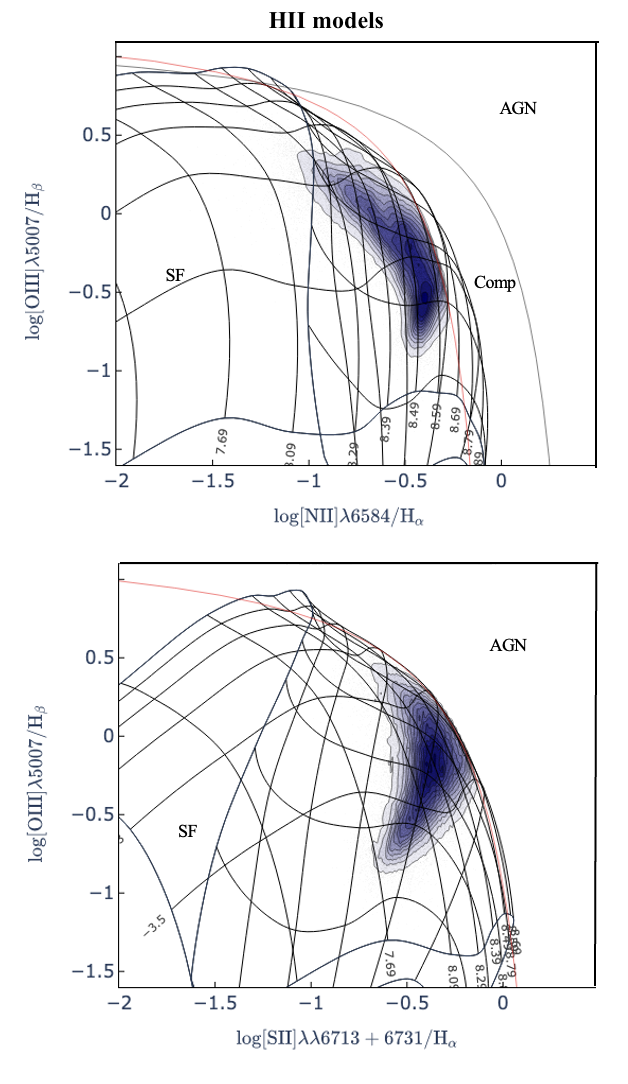}}%
    \caption{
     {\sc H ii} models (black lines) in the NII-BPT (\emph{top panel}) and in the SII-BPT diagram (\emph{bottom panel}).
    The {\sc H ii} models, generated with {\sc Cloudy}, have $\rm n_{\rm H} = 100 \ cm^{-3}$ and stellar ages $t_*$= 4 Myr. 
    Density curves, filled with different shades of purple, are drawn to show the distribution of the observed line ratios of the SF spaxels in the MaNGA  and GASP samples together. 
     Red lines are the \cite{kauffmann+2003} and \cite{kewley+2001} relations defining the SF regions in the NII-BPT and in the SII-BPT respectively. The black line in the NII-BPT is the \cite{kewley+2001} relationship which distinguishes Composite and Seyfert/LINER. The {\sc H ii} grids fold, due to the degeneracy between the metallicity and the ionization parameter, around 12 + log (O/H) = 8.6 - 8.7.}
    \label{fig:HIImodels}
\end{figure}
\begin{figure}
    \centering
    \makebox[0.2\textwidth]{
    \includegraphics[scale=0.45]{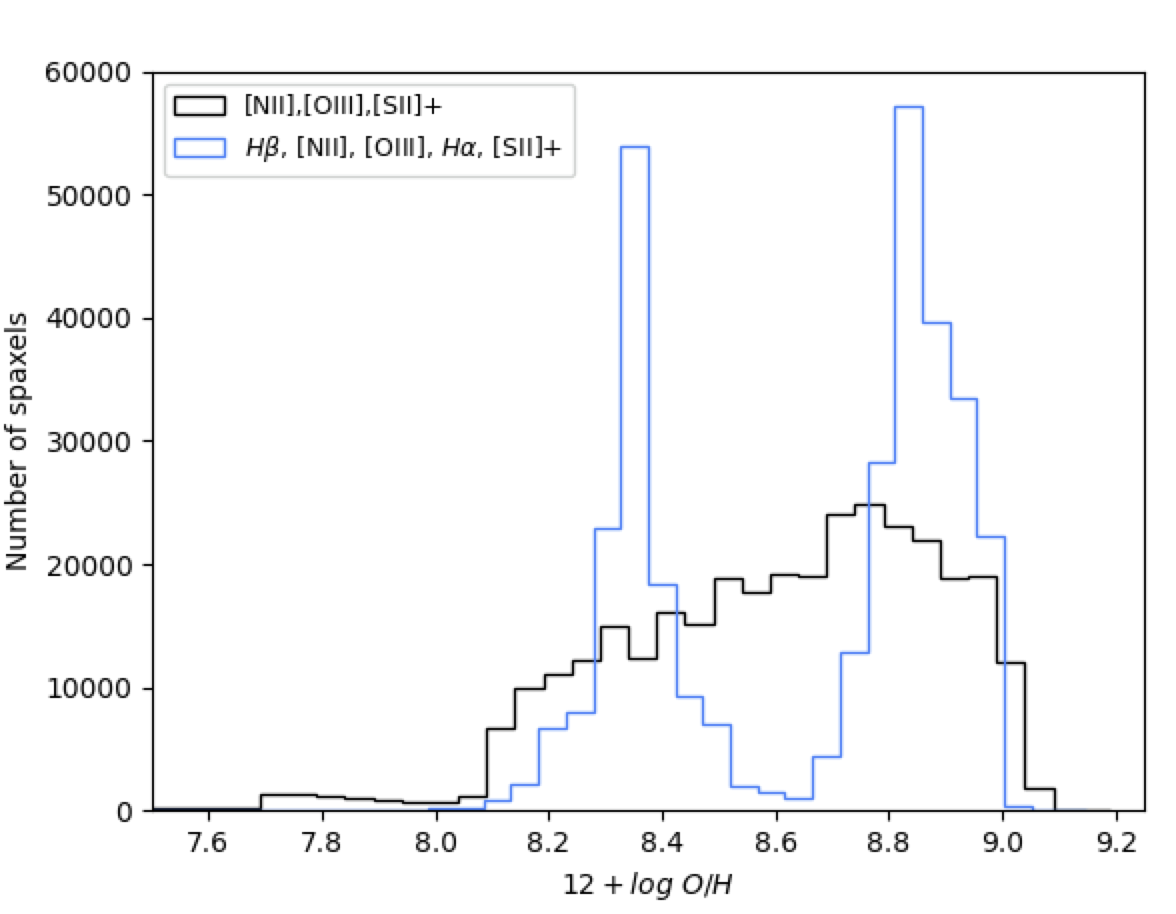}}%
    \caption{Histograms of the 12 + log (O/H) values inside all the spaxels classified as SF by the NII-BPT, in the MaNGA sample. The black histogram shows a uniform distribution in metallicity, obtained when using the \oiii~ and \nii~ lines, normalized for \sii. The blue histogram shows a strong bimodality, with a gap around 12 + log (O/H) $\sim$ 8.6, and is obtained when normalizing with H$\beta$ the set of lines: H$\beta$, \nii, \oiii, H$\alpha$, \sii. The bimodality is caused by the $\log(U)$ - $\log Z$ degeneracy of the models observed in the NII-BPT shown in Figure \ref{fig:HIImodels} (see Section \ref{sec:method}).}
    \label{fig:hist-SF-ALL}
\end{figure}

Particular care is necessary when selecting the emission lines to use in a {\sc Nebulabayes} analysis \citep[see][for details]{thomas+2018a}.
Among all the emission lines covered by our observational samples we 
notice that some widely used combinations are particularly affected by the degeneracy between the ionization parameter and the metallicity \citep[][and diagnostics therein]{dopita+2013}.
In particular, Figure \ref{fig:HIImodels} shows the \nii~/H$\alpha$ versus \oiii~/H$\beta$ for our observational sample with overlaid our own photoionization models, showing that the models are correctly reproducing the observations.  However, we also note the well-known folding in the \nii~-BPT and \sii~-BPT, which in our models happens around 12 + log (O/H) = 8.6. It follows that,
by using the line ratios \oiii~/H$\alpha$, \nii~/H$\alpha$ and \sii~/H$\alpha$ to constrain the parameters, we obtain that {\sc {\sc Nebulabayes}} does not converge to a solution for metallicities around 12 + log (O/H) = 8.59 \citep[similarly to][]{mingozzi+2020}, as it is clear from the blue histogram in Figure \ref{fig:hist-SF-ALL}, where we show the results for the MaNGA galaxies. Instead, the choice of using the line ratios \oiii~/\sii~ and \nii~/\sii~ produces the smooth metallicity distribution shown in Figure \ref{fig:hist-SF-ALL} as a black histogram. 
We stress that in the latter case the Balmer lines (H$\alpha$, H$\beta$) are not used to constrain the parameter space, but only to estimate the extinction correction as described in detail in Section \S \ref{sec:data}.


As already emphasized by previous studies that used a modeling technique similar to ours \citep{thomas+2019, perez-diaz+2021, mingozzi+2020},
we conclude that the degeneracy could have been broken with the fundamental addition of the \oiii~$\lambda$4363 auroral line  and/or a constrain on the ionization parameter through known relations \citep[e.g.,][]{perez-montero+2014,diaz+1991} between $\log(U)$ and  12 + log (O/H) (see Appendix \ref{sec:appendix-b}).  
However, we remind the reader that in this work it was not possible to include the auroral line \oiii~$\lambda$4363, because it is too faint for MaNGA and outside the MUSE spectral range for GASP targets, and other lines used in relations to constrain $\log(U)$, such as the \oii~$\lambda$3727, \siii~$\lambda$9069 and \siii~$\lambda$9532 lines, that are outside the MUSE wavelength range at the target's redshift. 

In conclusion, by applying the method presented in this Section, we are able to estimate spatially-resolved maps of the parameters 12 + log (O/H) and $\log(U)$ in case of ionization by stars, an AGN or a mixed contribution of them using the {\sc H ii}, AGN and Composite models described in Section \ref{sec:models}.

\section{RESULTS AND DISCUSSION} \label{sec:results}

We estimate the metallicity of the galaxy’s central regions, from spatially-resolved maps of the oxygen abundance, inside an aperture that scales with the
galaxy’s mass (or mass-scaled aperture), in order
to address the following question:

\begin{itemize}
    \item[i)]  Does a relation between the (host galaxy) stellar mass and metallicity of galaxies with AGN activity exist in RPS galaxies?
\end{itemize}

To draw the mass-scaled aperture, we compute the
projected distances from the galaxy center, using its
structural parameters such as the inclination and position angle, and select the spaxels within a projected distance of 0.5 ${R_{\rm e}}$.


In case of AGN host galaxies, we compute the median value of all the 12 + log O/H inside the AGN, Composite and SF spaxels contained by the aperture, using the corresponding models (see Section \ref{sec:models}). 
In the case of SF galaxies without AGN activity, we discard the emission classified as Composite, which in general is present inside a low fraction of spaxels (i.e., $\sim $5\% in the SF GASP sample and $\sim$ 13\% in the SF MaNGA sample), since in this case we cannot assume that its origin is the mixed AGN+SF contribution implied to generate our Composite models. Then, we compute the metallicities inside the SF spaxels using the {\sc H ii} models.

By using the same approach, we also estimate the median metallicity inside a fixed aperture of radius $r \sim$ 1 kpc (always dominated by AGN emission in case of AGN hosts) to address another open question:

\begin{itemize}
    \item[ii)]  Does the AGN in RPS galaxies show signs of metal enrichment or metal decrement with respect to the same physical region at the center of star-forming galaxies of similar masses? 
\end{itemize}




To understand if the results depend on the RPS, we answer the same questions for the galaxies part of the control samples (SF-FS, AGN-FS) which are located in the field and are undisturbed by RP. The results in the field galaxies are interesting on their own as 
it is still highly debated in the literature whether an NLR metallicity - stellar mass relation exists \citep[e.g.,][]{thomas+2019, dors+2020a,perez-diaz+2021} and if AGN are more or less metal-enriched than star-forming regions \citep[e.g.,][]{perez-diaz+2021,armah+2023}, since discrepant results have been found even without making a distinction based on the galaxy's environments, as discussed in the Introduction.

\subsection{The effect of different extraction apertures on spatially-resolved metallicity maps}

\begin{figure*}[ht]
    \makebox[\textwidth]{
    \includegraphics[scale=1]{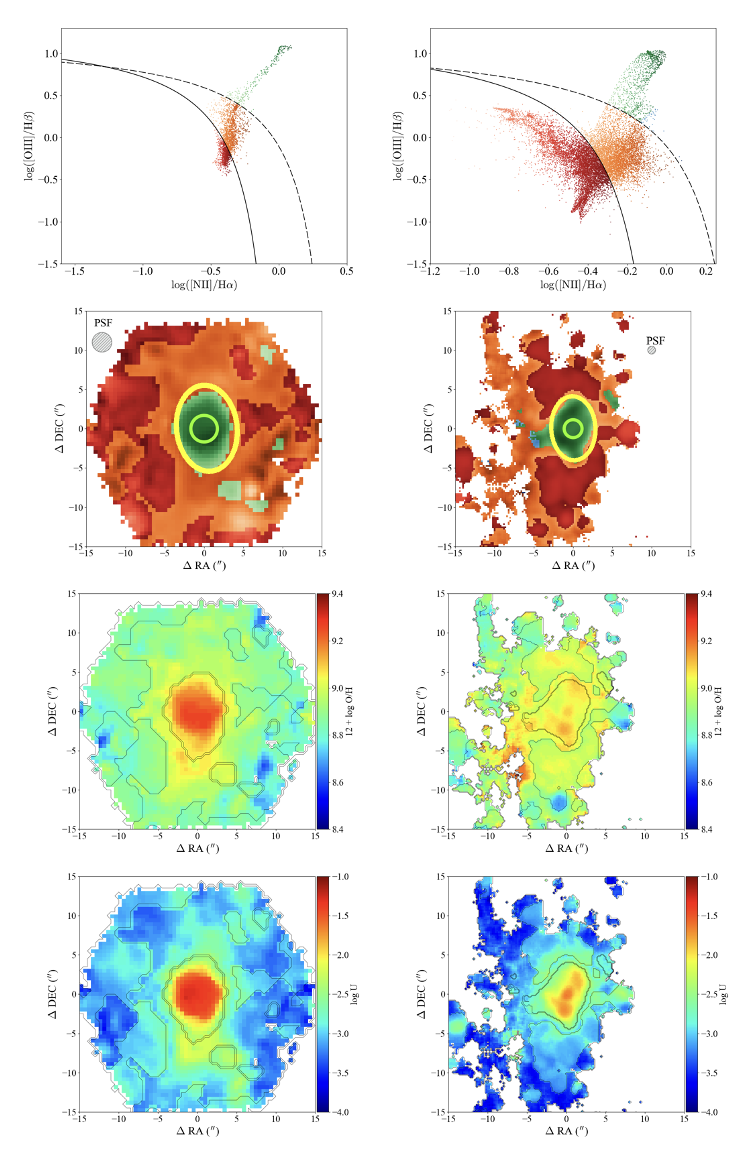}}%
    \vspace{-0.5cm}
    \caption{NII-BPT diagram and maps of the galaxy '8993-12705' (part of the AGN-FS, on the left) and JO201 (part of the AGN-RPS, on the right). (\emph{top panel}) NII-BPT diagram for all the spaxels in the galaxies, where in the case of JO201 we also include the spaxels of the stripped tail. SF spaxels are in red, Composite spaxels are in orange, LINER spaxels are in light blue and Seyfert spaxels are in green. Darker color shades indicate more intense line ratios, and viceversa. The black line is the \cite{kauffmann+2003} relation and the dotted black line is the \cite{kewley+2001} relation.  (\emph{middle panel}) Galaxy map color-coded according to the NII-BPT classification on which we draw the $r \sim$ 1 kpc (bright green circle) and $r \sim$ 0.5${R_{\rm e}}$ (yellow circle) apertures. 
   The $r \sim $1 kpc aperture is clearly dominated by AGN-only spaxels, while the $r \sim 0.5 R_{\rm e}$ includes a small fraction of SF/Composite spaxels. 
   The typical PSF size is shown in the top-left corner, with a grey circle.
   (\emph{bottom panels}) Galaxy map color-coded to the values of 12 + log (O/H) and $\log(U)$. The black contours, overlaid on the maps, divide regions classified as AGN/Composite/SF by the NII-BPT. The oxygen abundance 12 + log (O/H) varies between 8.4 and 9.2, the ionization parameter ranges between -4.0 and -1.0. }
    \label{fig:maps}
\end{figure*}

To illustrate how the choice of the aperture affects the AGN metallicity, we selected two galaxies from the GASP and MANGA samples, shown in Figure \ref{fig:maps}: both galaxies host Seyfert2-like nuclei according to the BPT and have similar stellar masses.
The top panel of Figure \ref{fig:maps} shows the NII-BPT diagrams of the field galaxy '8993-12705' ($z=0.030$, $\rm \log \ {\rm M}_*/{\rm M}_\odot$ = 10.96) and a zoom on the cluster galaxy JO201 ($z= 0.0446$, $\rm \log \ \rm M_*/\rm M_\odot$ = 10.79), which is experiencing strong RPS as discussed in \cite{poggianti+2017b}. 
The other panels of the same figure show the galaxy maps color-coded according to the NII-BPT classification, the metallicity and ionization parameter.
On the NII-BPT color-coded map, we overlay 
the yellow projected aperture extending up to 0.5 $R_{\rm e}$ 
and the green on-sky aperture extending up to 1 kpc from the galaxy center.
The 1 kpc aperture includes a higher or lower fraction of the galaxy's total light depending on the stellar mass, as opposed to the $r \sim$ 0.5 $R_{\rm e}$ aperture. 
However, the 1 kpc aperture has the advantage to include predominantly AGN spaxels, while
the 0.5$R_{\rm e}$ aperture in some galaxies
includes a non-negligible fraction of SF/Composite spaxels.
In this sense, the 1 kpc aperture is a better-suited choice to reduce the dependence of the AGN metallicity estimates on 
processes that are not linked to the presence of the AGN, as shown at the end of Section \S \ref{sec:results-2}.

To show the range of ionization parameter and metallicity spanned by the emission in the `nuclear regions' of our galaxy samples, in Appendix \ref{sec:appendix-a0} we show the \nii-BPT diagrams obtained with the emission line ratios within the extraction apertures $r < 0.5 \ R_{\rm e}$ and $r <$ 1 kpc color-coded according to the corresponding values of $\log (U)$ and 12 + log (O/H), for all the GASP and MaNGA galaxies.

Finally, we briefly comment on the metallicity and ionization parameter maps of the two galaxies shown in Figures \ref{fig:maps} (c) and (d). Further details on the metallicity profiles of our AGN sample will be discussed in a separate paper.

The galaxy '8993-12705' shows a strong inward increase in metallicity, which rapidly increases from 12 + log (O/H) $\sim$ 8.8 in the outer star-forming regions to 12 + log (O/H) $\sim$ 9.2 in the galaxy center. The transition from lower to higher metallicities is co-spatial with the increase of the AGN ionization parameter, which jumps from $ \log (U) \sim $ -2.5 to $\log (U) \sim$ -1.3. Star-forming regions show an average value of $\log (U) \ \sim$ -3.2 \citep[see also][]{thomas+2019}. On the other side, the metallicity in  JO201 peaks around 12 + log (O/H) $\sim$ 9.0 in the galaxy center in correspondence to the AGN, and we note that also the gas in the right lower side of the stripped tail shows similar values as well.  We also observe two high-metallicity and high-ionization parameter elongated regions symmetrically oriented with respect to the galaxy center. The peak of the AGN ionization parameter ($\log  (U) \sim - 1.6$) and of the metallicity (12 + log (O/H) $\sim$ 9.0) is presumably tracing  the actual position of the AGN, more precisely than the
NII-BPT classification map in which the AGN-like region is extending well beyond the NLR.

\subsection{Gas-phase metallicity of the AGN in RP stripped and undisturbed galaxies} \label{sec:results-1}

\begin{figure*}
    \centering
    \includegraphics[scale=0.6]{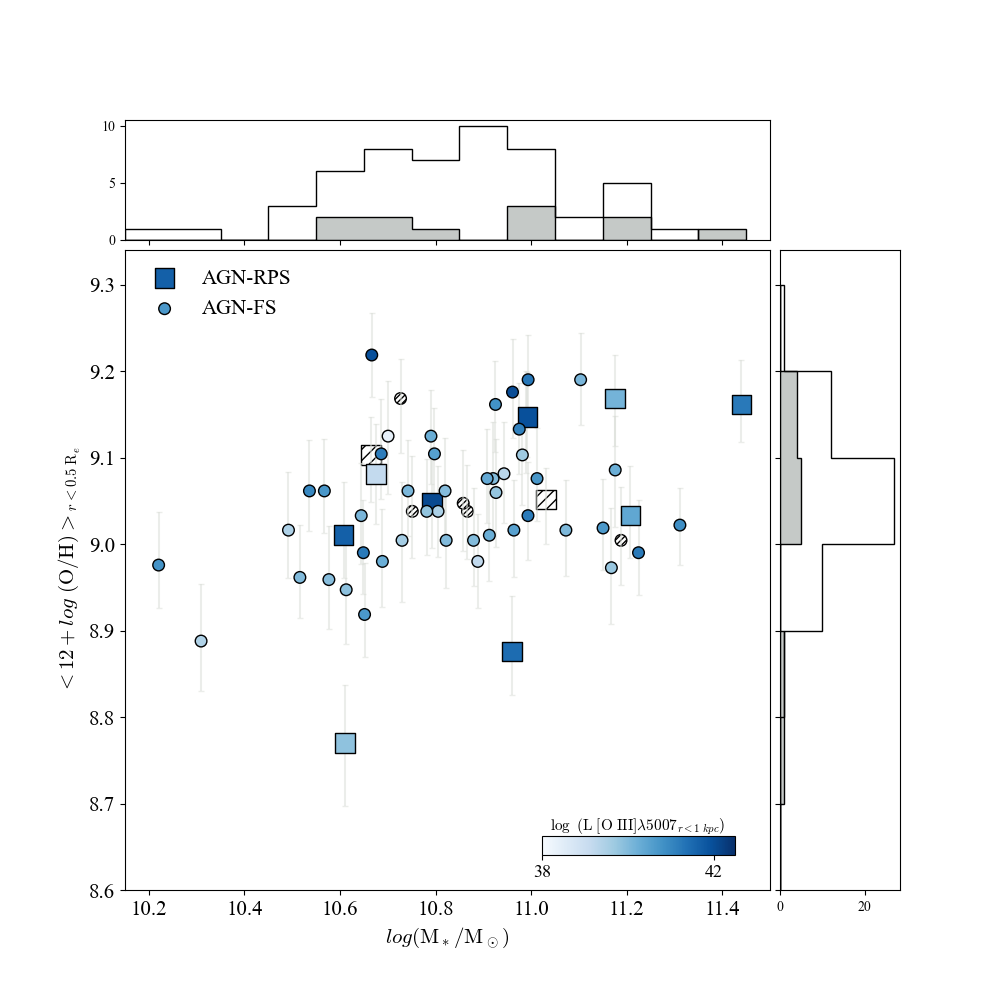}
    \caption{MZR of the AGN-RPS (squares) and AGN-FS (circles) color-coded according to their $L$\oiii ~. We show as dashed white symbols those AGN for which we could not estimate $L$\oiii ~ (see text for details). The 12 + log (O/H) is computed within the mass-scaled aperture ($r \sim$ 0.5 $R_{\rm e}$) and $L$\oiii ~ is computed within the fixed aperture ($r \sim 1$ kpc). 
    The error bars are the average values of the 16th and 84th percentiles of the PDF among all the spaxels within $0.5 R_{\rm e}$.
 The white and grey histograms (in the top and left insets) show the mass and metallicity distributions of the AGN-FS and the AGN-RPS. The two samples have similar ranges of oxygen abundances and  $L$\oiii ~.
 }
    \label{fig:agn-mzr}
\end{figure*}

Figure \ref{fig:agn-mzr} shows the metallicity as a function of the host galaxy stellar mass of the AGN-RPS (squares) and AGN-FS (circles) samples.
To compute metallicities, we consider the median value of 12 + log (O/H) in all the spaxels within $r \sim 0.5 \ R_{\rm e}$ from the galaxy center as a representative value for each galaxy.
We have verified that the mass-scaled aperture was always larger than the PSF (e.g., on-sky aperture with diameter $d \ \sim$ 2.5$\arcsec$ in MaNGA, and $d \ \sim$ 1$\arcsec$ in GASP), and therefore includes a well-resolved galactic region.
In support of the robustness of our results to a different choice of the extraction aperture,
Figure 3 in \cite{franchetto+2020} clearly shows that the mean (or median) value inside 0.5 $R_{\rm e}$ is consistent with the median values computed inside smaller apertures or at fixed galactocentric radii in our galaxies \citep[see also][]{moustakas+2006}.

Points in Figure \ref{fig:agn-mzr} are color-coded according to the integrated luminosity of the emission line \oiii~$\lambda$5007 (i.e., $L$\oiii~ hereafter) inside the fixed aperture of $r \sim 1$ kpc, which is a proxy of the bolometric luminosity of the central AGN \citep[e.g.,][]{berton+2015}. We calculated  $L$\oiii ~ only for  galaxies with at least 10 spaxels within $r \sim 1$ kpc powered by the AGN according to the BPT diagram, with S/N $>3$ in GASP or S/N $> 1.5$ in MaNGA for the lines listed in Section \S \ref{sec:data}. This selection restricts our sample to 9/11 AGN in GASP and 48/52 AGN in MaNGA. AGN with no reliable $L$\oiii ~ are shown as dashed white-colored symbols.
The $12 + \log $(O/H) and stellar mass distributions of the AGN-RPS and AGN-FS are shown as grey and white histograms, respectively, in the subpanels.

The two samples span the same range of $L$\oiii ~ and 12 + log (O/H), where the minimum values are $2.5 \times 10^{38} \ {\rm L}_\odot$ and  8.77, respectively, and  the maximum values are $1.2 \times 10^{42} \ {\rm L}_\odot$ and  9.22.

 A 2D Kolmogorov-Smirnov (KS) test could not exclude that the AGN-FS and AGN-RPS samples are drawn from the same parent distribution. This result suggests that the RPS is not playing a crucial role in shaping the metallicity within $r < 0.5 \ R_{\rm e}$  and the \oiii~ luminosity of the AGN ($r < 1$ kpc) in AGN hosts.

Galaxies of the AGN-RPS sample have slightly higher median  $L$\oiii ~ =  41.24$^{+ 0.65}_{-1.28}\ {\rm L}_\odot$ than the AGN-FS with median $L$\oiii ~ = 40.19$^{+0.88}_{-0.57}\ {\rm L}_\odot$. However, the values are consistent within the 16th and 84th percentiles of the $L$\oiii ~ and 12 + log (O/H) distributions. 
The higher $L$\oiii ~ luminosities of the AGN-RPS sample are presumably due to the preponderance of Seyfert-like nuclei in this sample.
In fact, the AGN-RPS has $\sim$ 50 \%  (6/11) of Seyfert 2 galaxies, while in the AGN-FS the Seyfert fraction is 16\% (9/53). 
 

Next, we study the relationship between the stellar mass and the AGN metallicity in the AGN-RPS and  AGN-FS samples, joined together. 
The Spearman correlation coefficient is R $\sim$ 0.27 with a p-value of 0.034, thus the test is not able to conclude that the two quantities are correlated. 
We argue that this can partially depend on the fact that the galaxies span a very limited range in stellar mass, as
 AGN are known to be located only in the most massive systems \citep[e.g.,][]{sanchez+2018,peluso+2022}. We also do not see a clear relationship between the stellar mass and $L$\oiii ~ from Figure \ref{fig:agn-mzr} and, accordingly to that, the Spearman test gives a correlation coefficient of R $\sim$ 0.25 with a p-value of 0.07.

\subsection{Comparison between metallicities of the nuclear regions in AGN and SF galaxies} \label{sec:results-2}

\begin{figure*}
    \centering
    \includegraphics[scale=0.9]{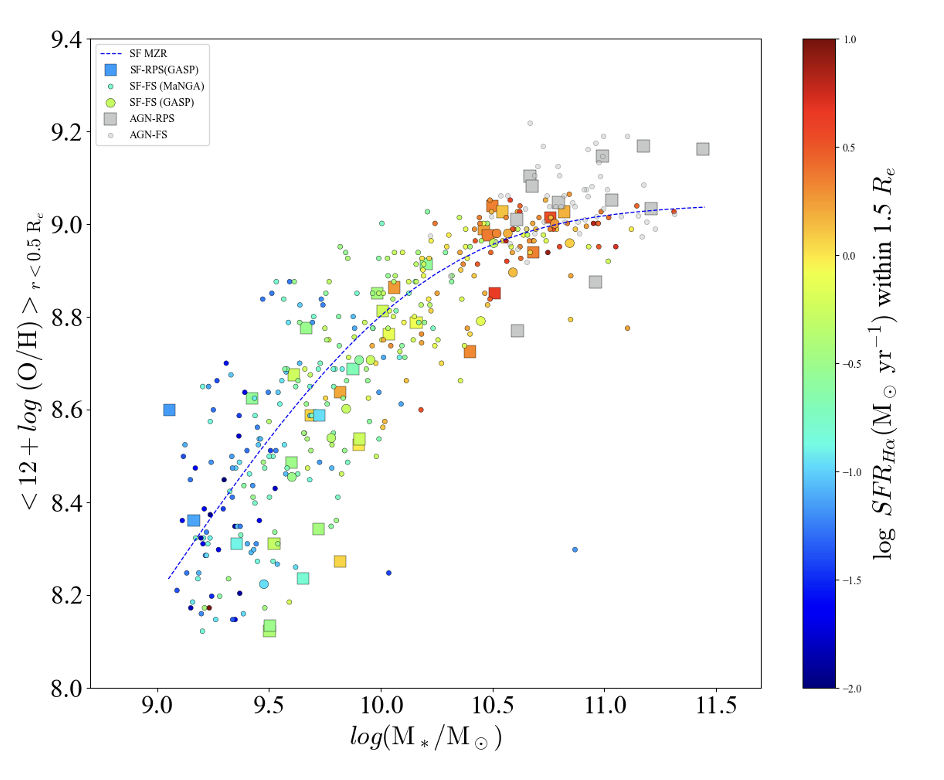}
    \caption{Mass-metallicity relation of the SF (colored points) and AGN (grey points) galaxies, with different symbols for the RPS (squares) and non-RPS (circles) samples. Metallicity is computed as the median value in all the spaxels (AGN/SF/Composite) within 0.5 $R_{\rm e}$. The blue dotted curve is the best fit for the SF galaxies.  SF galaxies are color-coded according to their SFR within 1.5 $R_{\rm e}$,  a proxy for the total SFR. 
    Overall,  AGN galaxies have higher metallicities than SF galaxies.
    } 
    \label{fig:mzr}
\end{figure*}

\begin{figure*}[htb]
    \centering
    \includegraphics[scale=0.9]{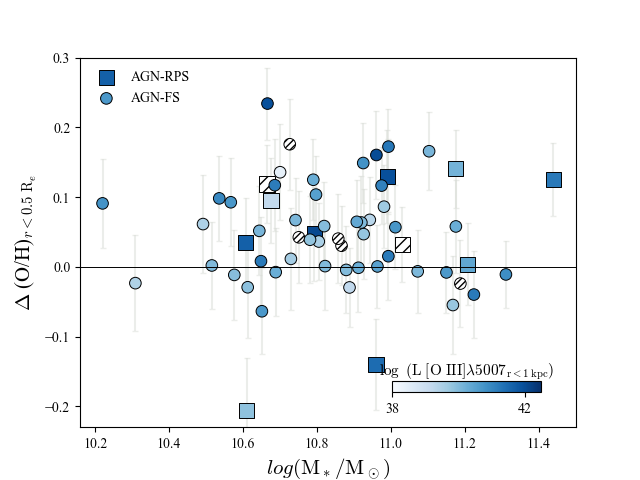}
    \caption{Residuals of the AGN-RPS (squares) and AGN-FS (circles) metallicity from the SF MZR, as a function of the galaxy stellar mass, color-coded according to $L$\oiii~ as in Figure \ref{fig:agn-mzr}. To compute the error bars on $\Delta$ (O/H), we consider the errors on the AGN metallicity and the errors on the SF MZR, computed as described in Section \S \ref{sec:results-2}. 
    The horizontal black solid line remarks the level of $\Delta$(O/H) = 0. AGN hosts show $\Delta$(O/H)$>0$ on average, except for 2/11 galaxies in the AGN-RPS sample that have  lower metallicity than SF galaxies.} 
    \label{fig:delta-oh-all}
\end{figure*}

To test previous literature findings \citep{armah+2023,thomas+2019}, we investigate the difference between the metallicity in the nuclear regions of AGN and SF galaxies. 
Even though the AGN-RPS and AGN-FS show the same MZ distributions (see Section \ref{sec:results-1}), in the first part of this section we still present the results separately for the two samples.


Figure \ref{fig:mzr} shows the MZR of the SF and AGN galaxies, with different symbols for RP-stripped (squares) and non-RP-stripped (circles) galaxies. The metallicity is computed as the median of all the values of 12 + log (O/H) within $r < 0.5 \ R_{\rm e}$.
The AGN galaxies are shown as grey symbols, while the SF galaxies are color-coded according to their SFR (H$\alpha)_{1.5Re}$, which is the SFR  within 1.5 $R_{\rm e}$ computed with the \cite{kennicutt+1998} relation, SFR (${\rm M}_\odot \rm yr^{-1}$) = 4.6 $\times 10^{-42} \ L_{\rm H\alpha}$ 
(erg s$^{-1}$), using the reddening-corrected H$\alpha$-flux.
This is the FoV of the MaNGA SF galaxies (see Section \S \ref{sec:manga-sample}), while for the GASP SF galaxies we computed the SFR (H$\alpha)_{1.5 R_{\rm e}}$ by excluding the spaxels beyond 1.5$R_{\rm e}$

To fit the mass-metallicity relation of star-forming galaxies (SF MZR),  shown in Figure \ref{fig:mzr} as the blue dotted line, we join the SF-RPS and SF-FS and we exploit the SF-FS galaxies to obtain the fit also at high stellar masses where the AGN are located.
In fact, the SF-RPS sample has only 3/37 galaxies with  $\log {\rm M}_* / {\rm M}_\odot > 10.5$, since (as seen in P22) the GASP-RPS AGN fraction is 51\% in the mass bin $\log \ ({\rm M}_* /{\rm M}_\odot) > 10$, while the SF-FS has 33/391 galaxies (i.e., 9\%) with $\log  \ ({\rm M}_*/{\rm M}_\odot) > 10.8$.

 It is worth noticing, though, that the SF-RPS show on average lower metallicities than the SF-FS, but the lowest metallicities of the SF-RPS are consistent with the scatter expected from the Fundamental MZR \citep{mannucci+2010} as discussed in detail in Appendix \ref{sec:appendix-c}.

To fit the SF MZR, we use the parametrized function from \cite{curti+2020} \citep[see also][]{mingozzi+2020}:

\begin{equation}
    \rm 12 + log (O/H) = Z_0 - \gamma / \beta \times \log \left[1 + \left(\frac{M}{{\rm M}_0}\right)^{-\beta}\right] 
   \label{eq:mzr-fit}
\end{equation}

where ${\rm Z}_0$ is the asymptotic value of metallicity at which the relation saturates, ${\rm M}_0$ is the characteristic turnover mass above which the metallicity asymptotically approaches the upper metallicity limit ($Z_0$) and $\beta$ quantifies how rapidly the curve approaches its saturation value. For ${\rm M}_* < {\rm M}_0$, the SF-MZR is  a power law of index $\gamma$. 
We fix the turnover mass M$_0$ = 10$^{10.1} {\rm M}_\odot$. To obtain the  best-fit parameters, we use the non-linear least squares (NLS) method which minimizes the residuals, weighted for the uncertainty on the datapoints ($\sigma_y$). $\sigma_{y}$ is the lowest value between $\sigma_{-}$ and $\sigma_+$, where $\sigma_{-}$ and $\sigma_+$ are the average values of the 16th and 84th percentiles of the metallicity PDF among all the spaxels within 0.5$R_{\rm e}$.

We obtain the following best-fit parameters: 
${\rm Z}_0 = 9.045 \pm 0.001$, $\gamma = 0.754 \pm 0.008$ and $\beta = 1.121 \pm	0.064$. The one standard deviation error on the parameter estimates is the squared variance (i.e., the diagonal) of the covariance matrix. 
We observe a plateau in the SF MZR at $\log ({\rm M}_* / {\rm M}_\odot) > 10.5$ \citep[similarly to][]{tremonti+2004} where the metallicity is $<$12 + log (O/H)$>_{0.5 \ R_{\rm e}}$ $\sim$ 9.0 independently of the stellar mass.

Figure \ref{fig:delta-oh-all} shows the residuals of the AGN metallicities from the SF MZR, $\Delta$ (O/H)$_{r < 0.5 \ R_{\rm e}}$, which is the difference between the metallicity of the AGN and the one computed with equation (\ref{eq:mzr-fit}). AGN hosts predominantly lie  above the  SF-MZR, suggesting that the presence of the AGN is enhancing the oxygen abundance in the galactic nuclei. As expected from the results presented in Section \S \ref{sec:results-2}, this result is independent of the presence of RPS, since 
the AGN-RPS sample shows a similar enhancement in metallicity as the AGN-FS sample.
Overall, we find that the median offset of the combined AGN sample (RPS and FS)  from the SF MZR is $\Delta$ (O/H) $_{r<0.5 \ R_{\rm e}}$ =  0.047 dex, which is consistent with previous findings \citep[e.g.,][]{thomas+2019}. 

Interestingly, two galaxies (JO206 and JO171) from the AGN-RPS show a  metallicity that is lower than that found in SF galaxies, which are the outliers in Figure \ref{fig:delta-oh-all}. 
The AGN with the lowest metallicity, JO171, is a very peculiar object as it is a Hoag-like post-merger whose central metallicity is not directly linked with the total mass \citep{moretti+2018}. While for the galaxy JO206 the interpretation is less clear and would require further analysis, for example by exploring the possible presence of metal-poor inflows of gas \cite[as recently seen in e.g.,][in the IR regime]{perez-diaz+2023} or a particularly strong AGN feedback \citep[][]{armah+2023}.

\begin{table*}[h!t]
\begin{tabular}{p{2.5cm} p{4cm} p{4cm} p{3cm}}
\midrule  
$\log ({\rm M}_{0}/{\rm M}_\odot)$ & (12+ log O/H)$_{{\rm AGN,1 kpc}}$ & (12+ log O/H)$ _{{\rm SF,1 kpc}}$  & $\Delta$ (O/H)$_{r<1 \rm kpc}$\\
\midrule
10.50  & $9.069_{-0.045}^{+0.152}$ &  $8.989_{-0.045}^{+0.152}$ & $0.080_{-0.055}^{+0.065}$ \\
10.70  & 9.033$_{-0.045}^{+0.152}$ &  8.989$_{-0.045}^{+0.152}$ & 0.044$_{-0.059}^{+0.108}$ \\
10.90  & 9.119$^{+0.152} _{-0.045}$ &  9.014$^{+0.152} _{-0.045}$ & 0.104$^{+0.059} _{-0.113}$ \\
11.10  & 9.069$_{-0.045}^{+0.152}$ &  9.014$_{-0.045}^{+0.152}$ & 0.054$_{-0.046}^{+0.153}$ \\ 
\midrule
\end{tabular}
\caption{Columns: 1) central mass of the mass bins ($\log {\rm M}_{0}/{\rm M}_\odot$) in which there are more than 5 AGN galaxies; 2,3) median metallicities of the AGN  and SF galaxies inside the mass bin, with the 16th/84th percentiles of the distribution (12+ log O/H$_{{\rm AGN,1 kpc}}$ and 12+ log O/H$\rm _{SF,1kpc}$ respectively); (4) values of $\Delta$ (O/H)$_{r<1 \rm kpc}$ \, which are obtained as the difference between  (12+ log O/H)$_{{\rm AGN,1 kpc}}$  and  (12+ log O/H)$\rm _{SF,1 kpc}$; the errors are computed propagating the errors on (12+ log O/H)$_{{\rm AGN,1 kpc}}$ and (12+ log O/H)$_{ {\rm SF,1 kpc}}$.}
\label{table:fixed_mass}
\end{table*}

\begin{figure*}[htb]
    \makebox[\textwidth]{
    \includegraphics[scale=0.6]{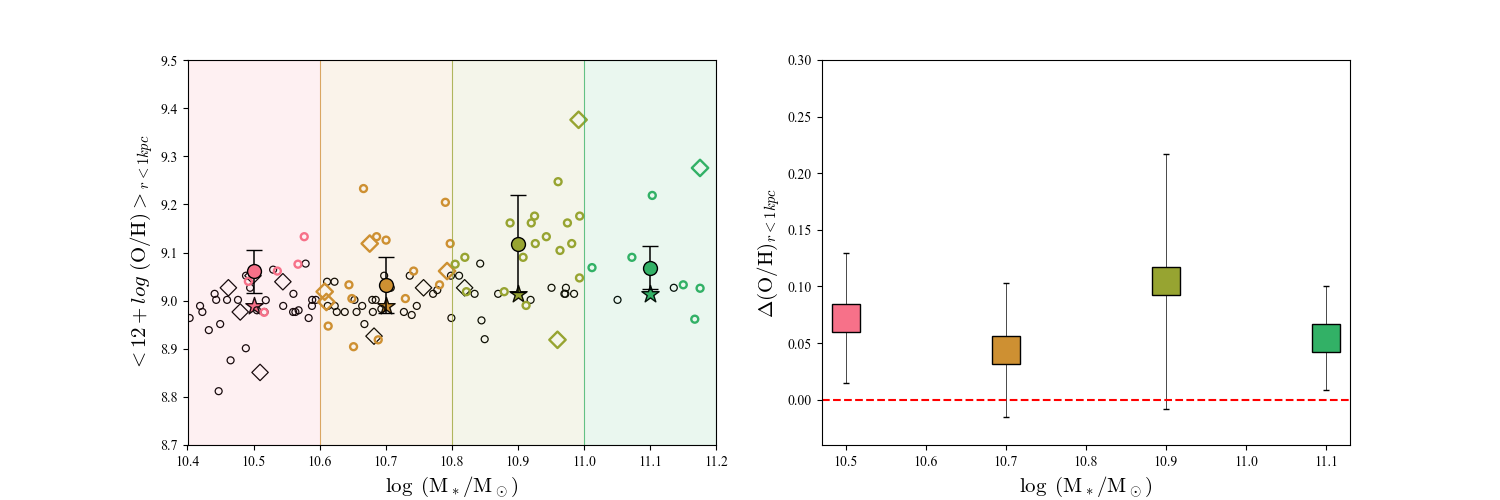}}%
    \caption{(left panel) Mass-metallicity relation of the AGN-RPS galaxies (edge-colored squares),  AGN-FS galaxies (edge-colored circles), SF-RPS galaxies (edge-black squares) and SF-FS galaxies (edge-black circles) with stellar masses $\log ({\rm M}_*/{\rm M}_\odot) \geq 10.4$, where the metallicities are the median values of 12 + log (O/H) within 1 kpc from the galaxy centers. 
    The filled-colored circles are the median metallicity  (12 + log (O/H)$_{AGN,1 kpc}$) of the AGN inside the mass bin (i.e., strips of different colors) with the errors given by the 16th/84th percentile of the distribution. The filled-colored stars are the median metallicities (12 + log (O/H)$_{SF,1 kpc}$) of the SF galaxies inside the mass bin. 
    (right panel) Difference between (12 + log (O/H)$_{AGN,1kpc}$) and (12 + log (O/H)$_{SF,1kpc}$) as a function of  stellar mass. 
    Within the same physical region, galaxies hosting AGN are more enriched in metals than those without AGN activity.
    }
    \label{fig:fixed-mass}
\end{figure*}

To have an estimate of the AGN metallicity without a significant contribution from gas ionized by stars, 
we also computed the metallicities inside the fixed aperture of radius $r \sim$ 1 kpc from the galaxy center, which is always dominated by the emission from Seyfert/LINER-classified spaxels in case of AGN hosts.

Figure \ref{fig:fixed-mass} (left panel) shows the metallicities within 1 kpc for galaxies with $\log ({\rm M}_*/{\rm M}_\odot)>10.4$.
We consider separately galaxies in stellar mass bins of 0.2 dex width (i.e., stripes of different colors), since we want to avoid the dependence of the metallicity estimates on the portion of the galaxy covered by the fixed aperture, which changes with the stellar mass. 
 
We compute the median 12 + log (O/H) of AGN metallicities inside each mass bin (12+log O/H$\rm _{AGN,1 kpc}$, filled-colored circles) only when there are more than five AGN galaxies inside that bin.


  
Qualitatively, results do not change depending on the chosen aperture
and, as for the mass-scaled aperture, the AGN galaxies show higher metallicities than SF galaxies. 
 Figure \ref{fig:fixed-mass} (right panel) shows $\Delta$(O/H)$_{1 kpc}$ which is the difference between the metallicities of the AGN (12+log (O/H)$\rm _{AGN,1 kpc}$) and SF (12+$\log$ (O/H)$\rm _{SF,1 kpc}$) galaxies with similar stellar masses, which basically quantifies how much the NLR is enriched in metals with respect to a region with the same physical extension but at the center of star-forming galaxies. 
 The red dotted line remarks the level at which $\Delta$(O/H)$_{1 kpc} = 0$. 
In Table \ref{table:fixed_mass} we list the central mass of the bin, (12+log O/H)$\rm _{SF,1 kpc}$, (12+log O/H)$\rm _{AGN,1 kpc}$ and $\Delta$(O/H)$_{r<1 \rm kpc}$ in each mass bin. 
The errors on $\Delta$(O/H)$_{r<1 kpc}$ are calculated considering the errors on (12+log O/H)$\rm _{AGN,1 kpc}$ and (12+log O/H)$\rm _{SF,1 kpc}$.
The offset $\Delta$(O/H)$_{r<1 \rm kpc}$ is positive in each bin of mass and ranges between 0.044 dex and 0.065 dex depending on the stellar mass, which is consistent within the errors with the offset of 0.06 dex measured by \cite{thomas+2019}. The aperture used by \cite{thomas+2019} to integrate the metallicity  
is comparable in extension with the fixed aperture of 1 kpc at our targets' redshift, as discussed in detail in the following sections.

\subsection{Comparison with the literature}
By using a similar approach to ours, \cite{perez-diaz+2021} find that AGN (both Seyferts and LINERs) galaxies do not follow a mass-metallicity relation and that Seyfert 2 have slightly higher chemical abundances than SF galaxies, in the mass range $ 9 \leq \log \ ({\rm M}_*/{\rm M}_\odot) \leq 12$.
However, they also find that LINER galaxies have lower abundances than SF galaxies. 
\cite{perez-diaz+2021} 
use Bayesian inference to compare {\sc Cloudy} v17.01 models and observations by exploiting the code HCM \citep[][]{perez-montero+2014,perez-montero+2019} in a sample of 143 SF, LINER and Seyfert galaxies observed with the Palomar Spectroscopic Survey. One of the main differences with our analysis is that \cite{perez-diaz+2021} consider galaxies independently from their environments. 
On the contrary, we consider here the effects of AGN in determining the metallicity of their host galaxy in the dense cluster environment, even if our sample is biased towards those showing optical signatures of RPS. The field sample of galaxies is, instead, complete.
Being aware of that, we find a consistent offset between SF and AGN metallicities to that found in \cite{thomas+2019},  which uses the code  {\sc Nebulabayes} and SDSS data to compute the MZR in a sample of 7,669 Seyfert 2 galaxies and 231,429 SF galaxies. 
They also find that the active galaxies follow a mass-metallicity relation in the mass range $10.1 \leq \log \ {\rm M}_*/{\rm M}_\odot \leq 11.3$ since the nuclear metallicity in Sy2s increases of $\sim$ 0.1 dex over a stellar mass range of 1.3 dex. It is worth noticing, though, that the value 0.1 dex is of the same order than the errors on the metallicity estimates derived with {\sc Nebulabayes} (see e.g., Table \ref{table:fixed_mass} of this paper).
The offset of the oxygen abundance in Sy2s  with respect to the MZR of the star-forming galaxies is $\sim$ 0.09 dex, but 
reduces to $\sim$ 0.06 dex when considering the contribution to the offset coming from the fact that the metallicity in the Seyfert 2 and star-forming samples was constrained using different emission lines. 
The scatter of 0.06 dex is consistent (within the error bars) with the scatter measured in this work using the $r \sim$ 1 kpc aperture (i.e., 2 kpc in diameter), which indeed ranges between 0.04 and 0.07 dex. Our fixed aperture has a diameter of $\sim$ 2.5$\arcsec$ at our target's redshifts, which is fairly similar to the  Sloan fiber's diameter of 3$\arcsec$ used by \cite{thomas+2019}.


Nonetheless, other works find opposite results. \cite{armah+2023} find lower values of 12 + log (O/H) abundances (with a mean difference of 0.2-0.5 dex) in AGN hosts than in SF galaxies, using an unbiased sample of Seyfert nuclei in the local universe ($z \leq$ 0.31) from the BAT AGN Spectroscopic Survey \citep[BASS,][]{oh+2022}, which select AGN from their hard X-ray band emission (14-195 keV). These authors compute the AGN metallicities using 
the \cite{carvalho+2020} and \cite{storchi-bergmann+1998} calibrators, based on photoionization models. By using a similar approach, \cite{donascimento+2022} study the metallicity profiles of a sample of 107 Seyfert galaxies using the spatially resolved data from the SDSS-IV MaNGA and the \cite{carvalho+2020} and \cite{storchi-bergmann+1998} calibrators.
They compute the integrated AGN metallicity within the central 2.5$\arcsec$ and compare it with the value extrapolated from the radial oxygen abundance profile of {\sc H ii} regions in the galaxy disc. The oxygen abundance in the {\sc H ii}  regions is obtained with the calibrator from \cite{perez-contini+2009}. 
We find 9 AGN galaxies in common with the \cite{donascimento+2022}'s sample (which is indeed the number of Seyfert galaxies in our AGN-FS drawn from the MaNGA survey, see also Section \S \ref{sec:results-1}). 
We measure the integrated metallicity inside an on-sky aperture of 2.5$\arcsec$ centered on the galaxy, as in \cite{donascimento+2022}, but using the metallicity maps obtained in this work.
We find that the median difference between the 12 + log (O/H) measured in \cite{donascimento+2022} and ours is $-0.4$ dex when considering their estimates with the \cite{carvalho+2020} (C20) calibrator, and $-0.43$ dex when considering their oxygen abundances computed with \cite{storchi-bergmann+1998} (SB98). This is larger than the average difference between the NLR metallicity and the extrapolated value found by the authors, which ranges between  0.16 to 0.30 dex.
By estimating the metallicities with the SB98 and C20 calibrators inside all the AGN spaxels in our MaNGA and GASP samples, we find that the values of 12 + log(O/H) computed with our method and with these other calibrators are well-correlated with each other (i.e., $r=0.48$ in case of the 12 + log(O/H)$_{\rm SB98}$ and $r=0.57$ in case of the 12 + log (O/H)$_{\rm C20}$). However, following the approach of \cite{perez-diaz+2021},
we find an offset of 0.387 dex with RMSE = 0.12 dex between the 12 + log (O/H)$_{\rm C20}$ and 12 + log (O/H)$_{{\rm Nebulabayes}}$, while we find an offset of 0.391 dex with RMSE = 0.11 dex between the 12 + log (O/H)$_{\rm SB98}$ and 12 + log (O/H)$_{{\rm Nebulabayes}}$.
Therefore, we conclude that the \cite{carvalho+2020} and \cite{storchi-bergmann+1998} calibrators give systematically lower values of metallicity than the method applied throughout this work and this is the reason for the discrepancy between our findings and those in \cite{donascimento+2022}. 
We stress that even higher offsets are found in the literature when comparing different methods: for example, the offset found by \cite{perez-diaz+2021} when
comparing their method with the code {\sc NebulaBayes}  (but coupled
with the {\sc Mappings}  models, instead of the {\sc Cloudy}  models adopted by us) is 0.8 dex.
\cite{perez-diaz+2021} attributed this offset to the different power-law slope adopted in the {\sc Mappings} and their models ( $\alpha$ = -2.0 and  $\alpha$ = -0.8 respectively). However, in our case  $\alpha$ = -2.0 produces a significantly lower offset.
We therefore conclude that a detailed treatment of the possible effects that lead to these discrepancies involves a complex combination of the assumptions underlying each model, whose discussion is beyond the scope of this paper.


\section{SUMMARY}
\label{sec:discussion}
 In this paper, we have investigated the effect of RPS on the AGN metallicity of 11 Seyfert/LINER galaxies, by comparing their mass-metallicity distribution with that of 52 Seyfert/LINER galaxies undisturbed by RP. 
 We also studied the impact of the presence of a central AGN on the metal content of galactic nuclei, both in case of RPS and not, by exploring the difference between the metallicity at the center of AGN and SF galaxies.
To do so, we exploit IFU data from the GASP and MaNGA surveys, and we measure their metallicities using the {\sc Nebulabayes} code and a set of 
AGN, Composite and {\sc H ii} photoionization models generated with the version of the code {\sc Cloudy} v17.02.

Our main findings are summarized as follows:

\begin{itemize}
    \item AGN galaxies either experiencing RPS or not generally have the same distribution in the mass-metallicity diagram and span the same range of $L$\oiii ~ luminosity. This result suggests that the stripping is not impacting significantly the integrated metallicity and \oiii~ luminosity of the central AGN, at least when looking at a relatively large sample of galaxies;

    \item The AGN-RPS and AGN-FS galaxies do not seem to follow a mass-metallicity relation, as shown in Figure \ref{fig:agn-mzr}, within the short range of stellar masses they cover.
    
    \item Thanks to the use of IFU data, we were able to test our results by integrating the metallicities inside different extraction apertures. Independently from the extraction aperture and the RPS, AGN galaxies show on average enhanced metallicity with respect to SF galaxies at fixed stellar mass. 
    The difference between the metallicity at the centers of AGN and SF galaxies reaches values up to 0.2 dex when using the aperture with $r \sim 0.5 \ R_{\rm e}$, while the median difference between metallicities computed with the 1 kpc aperture ranges from 0.04 dex to 0.07 dex, depending on the host galaxy's stellar mass. 
\end{itemize}

In summary, our results show that the presence of the AGN implies higher metallicities in the nuclei of galaxies but that the RPS is not playing a role in changing either the AGN metallicity or \oiii~ luminosity. 
\\
\\

\section{Acknowledgements}
Based on observations collected at the European Organization for Astronomical Research in the Southern Hemisphere under ESO program 196.B-0578. This project has received funding from the European Research Council (ERC) under the European Union's Horizon 2020 research and innovation program (grant agreement No. 833824). We acknowledge financial contribution from the grant PRIN MIUR 2017 n.20173ML3WW\_001 (PI Cimatti).
We thank the anonymous referee for their comments that have helped us to improve the paper.
This work made use of the {\sc KUBEVIZ} software which is publicly available at \url{https://github.com/matteofox/kubeviz/}. The development of the KUBEVIZ code was supported by the Deutsche Forschungsgemeinschaft via Project IDs: WI3871/1-1 and WI3871/1-2.

This work makes use of data from SDSS-IV. Funding for SDSS has been provided by the Alfred P. Sloan Foundation and Participating Institutions. Additional funding toward SDSS-IV has been provided by the U.S. Department of Energy Office of Science. SDSS-IV acknowledges support and resources from the Center for High-Performance Computing at the University of Utah. The SDSS website is www.sdss.org. This research made use of Marvin, a core Python package and web framework for MaNGA data, developed by Brian Cherinka, José Sánchez- Gallego, and Brett Andrews \citep{cherinka+2019}. SDSS-IV is managed by the Astrophysical Research Consortium for the Participating Institutions of the SDSS Collaboration including the Brazilian Participation Group, the Carnegie Institution for Science, Carnegie Mellon University, the Chilean Participation Group, the French Participation Group, Harvard-Smithsonian Center for Astrophysics, Instituto de Astrofísica de Canarias, The Johns Hopkins University, Kavli Institute for the Physics and Mathematics of the Universe (IPMU)/University of Tokyo, Lawrence Berkeley National Laboratory, Leibniz Institut für Astrophysik Potsdam (AIP), Max-Planck-Institufür Astronomie (MPIA Heidelberg), Max-Planck-Institut für Astrophysik (MPA Garching), Max-Planck-Institut für Extra- terrestrische Physik (MPE), National Astronomical Observa- tory of China, New Mexico State University, New York University, University of Notre Dame, Observatário Nacional/ MCTI, The Ohio State University, Pennsylvania State University, Shanghai Astronomical Observatory, United King- dom Participation Group, Universidad Nacional Autónoma de México, University of Arizona, University of Colorado Boulder, University of Oxford, University of Portsmouth, University of Utah, University of Virginia, University of Washington, University of Wisconsin, Vanderbilt University, and Yale University.
The MaNGA data used in this work are publicly available at \url{http://www.sdss.org/dr15/manga/manga-data/}.

\bibliography{arxiv}{}
\bibliographystyle{aasjournal}

\clearpage
\appendix

\section{The lack of {\sc H ii} models with \nii/H$\alpha$ and \sii/H$\alpha$ ratios close to BPT demarcation lines} \label{sec:appendix-a}

\begin{figure*}[ht]
  \makebox[\textwidth]{
    \includegraphics[scale=0.6]{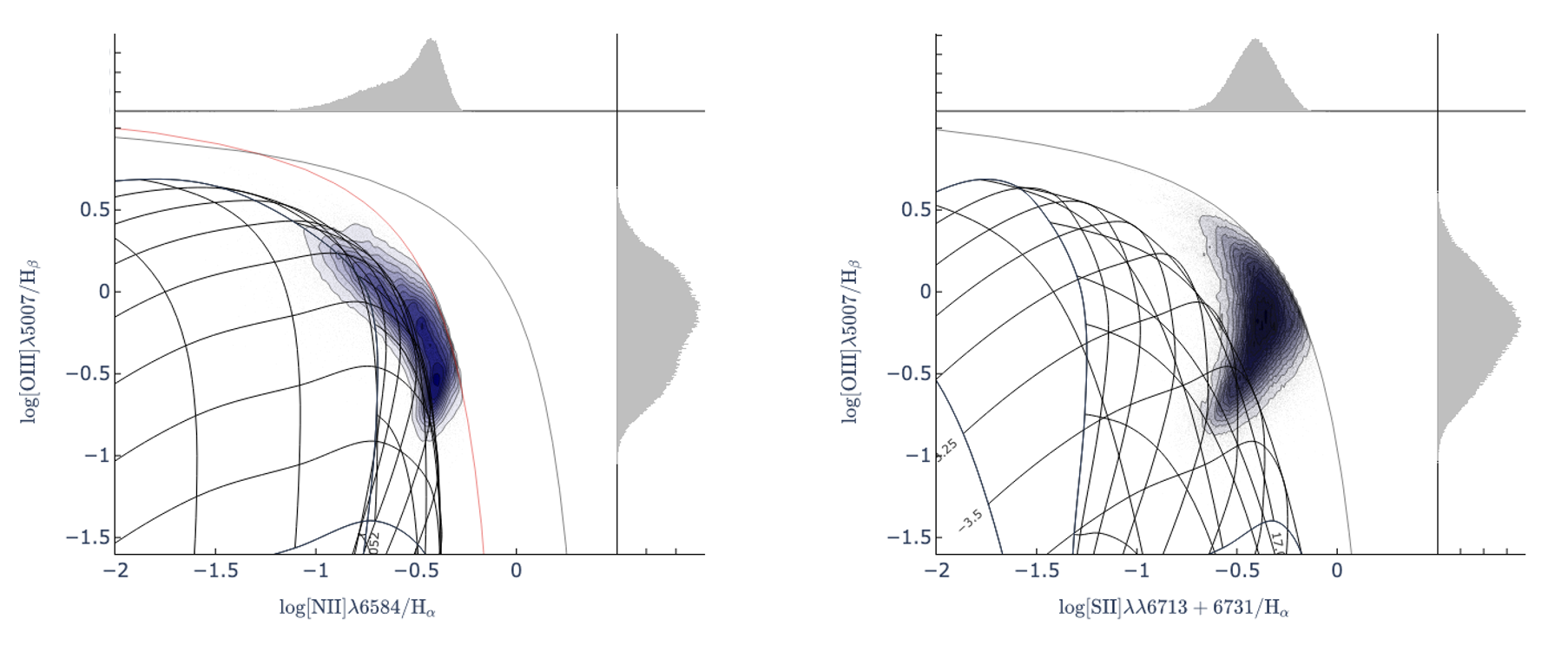}}%
    \caption{NII- and SII- BPT diagrams with overlaid the {\sc H ii} models (black lines) provided with the code {\sc Nebulabayes}. The model's gas pressure is log P/k (cm$^{-3}$/K)  = $5$. The red line is the empirical Kauffmann+2003 relation, while the black lines are the theoretical \cite{kewley+2001} relationships. The distribution of the observed star-forming line ratios (from GASP and MANGA galaxies) is shown by density curves filled with shades of purples, where darker colors indicate higher densities.}
    \label{fig:ALL-grids}
\end{figure*}

{\sc Nebulabayes} is provided with models generated with the code Mappings V \citep{sutherland-dopita+2017}  in case of ionization from both stars and AGN \citep[][ T18 models hereafter]{thomas+2018a}. 
However, Figure \ref{fig:ALL-grids} shows clearly the lack of {\sc H ii} models with \nii/H$\alpha$ and \sii/H$\alpha$ line ratios close to the BPT demarcation lines, which are the \cite{kauffmann+2003} empirical relationship in the NII-BPT
and the theoretical \cite{kewley+2001} relationship in the SII-BPT \citep[see][for a review of these demarcation lines]{law+2021}.
The discrepancy between models and observations  is consistent with that found in {\sc H ii} models generated with the previous version of the code Mappings IV \citep[][D13 models]{nicholls+2012,dopita+2013}. The {\sc H ii} models described in \cite{kewley+2019b} 
partly recover models with generally higher \nii/H$\alpha$ and \sii/H$\alpha$ than T18 and DR13, 
but still too weak to approach the line ratios of the \cite{kauffmann+2003} and \cite{kewley+2001} relations \citep[Fig. 11 in][]{kewley+2019a}. 
 \cite{byler+2017} generate {\sc H ii} models with the code {\sc Cloudy} v13 which come much closer to the BPT demarcation lines in both the NII- and SII- BPT diagrams, but only the use of the version of the code {\sc Cloudy}
 v17 \citep[][]{ferland+2017} leads to model predictions that are capable to follow exactly the \sii/H$\alpha$ and \nii/H$\alpha$ line ratios of the demarcation relations \citep{law+2021, belfiore+2022}.
Because of it, we generate {\sc H ii} models with the code {\sc Cloudy} v17 as described in Section \S \ref{sec:models} based on the prescription presented in \cite{byler+2017} \citep[see][for a similar approach] {belfiore+2022}. 

\section{CHOICE OF LINES TO USE IN NEBULABAYES} \label{sec:appendix-b}

In Section \S \ref{sec:method}, we have shown that we are unable to estimate metallicities around the value 12 + log (O/H) $\sim$ 8.6 by exploiting the H$\beta$-normalized lines: \oiii~$\lambda$5007, H$\alpha$, \nii~$\lambda$6584, \sii~$\lambda$6716 and \sii~$\lambda$6731. We interpret this result as a consequence of the folding of the {\sc H ii} models around 12 + log (O/H) $\sim$ 8.6 in the BPT diagrams caused by the well-known degeneracy \citep{dopita+2013} between $\log(U)$ and 12 + log (O/H). 
We summarize here those works using a similar approach to ours to compute the metallicities but without finding signs of degeneracy.



\cite{thomas+2019} use the code {\sc Nebulabayes} and the H$\beta$-normalized strong emission lines listed above, with the addition of the lines: \oii~$\lambda\lambda$3726/29, \neiii$\lambda$3869, \oiii~$\lambda$4363 and He I $\lambda$5876. 



Similarly, \cite{perez-diaz+2021} use the Bayesian code HCM \citep{perez-montero+2014,perez-montero+2019}, give in input the same set of lines listed above, with the only addition of the auroral line \oiii~$\lambda$4363 (if measured) and normalize them with H$\beta$. 

The code HCM assumes a relation between 12 + log (O/H) and $\log(U)$ \citep[from][]{perez-montero+2014} when the auroral line \oiii~$\lambda$4363 is not measured, as predicted 12 + log (O/H)'s values are not valid if the \oiii~$\lambda$4363 is not included \citep[see Figure 2 in][]{perez-montero+2014}.



\cite{mingozzi+2020} use the lines included in this work, with the only addition of the \oii doublet and the \siii~ $\lambda\lambda$9069,9532,  and an updated version of the code IZI \footnote{https://github.com/francbelf/python\_izi} \citep{blanc+2015}, from which {\sc Nebulabayes} was developed. The authors observe a bimodal metallicity distribution, peaking at 12 + log(O/H) $\sim$ 9 and at 12 + log(O/H) $\sim$ 8.6, with a gap around 12 + log(O/H) $\sim$ 8.8. They discuss how the bimodality is probably caused by the degeneracy between 12 + log(O/H) and log(q), and constrain the ionization parameter with the \siii~$\lambda\lambda$9069,9532 lines, using the \cite{diaz+1991} relation between the \siii/\sii  and log(q) to set a Gaussian prior on log(q). After this adjustment, the bimodality disappears. 
We conclude that a relationship \citep[e.g][]{diaz+1991, perez-montero+2014} between the ionization parameter and the metallicity, or the use of auroral lines (such as \oiii~$\lambda$4363), is necessary to measure 12 + log (O/H) when exploiting line ratios normalized by H$\beta$ line. 
We also argue more generally that line ratios involving Balmer lines (e.g., H$\beta$, H$\alpha$) are powerful tools to distinguish between star formation and AGN ionization (e.g., BPT diagrams), at the cost to be strongly degenerate in terms of the ionization parameter and not the better-suited choice to estimate the metallicity.

\begin{figure}[]
    \centering
    \makebox[\textwidth]{
    \includegraphics[scale=0.37]{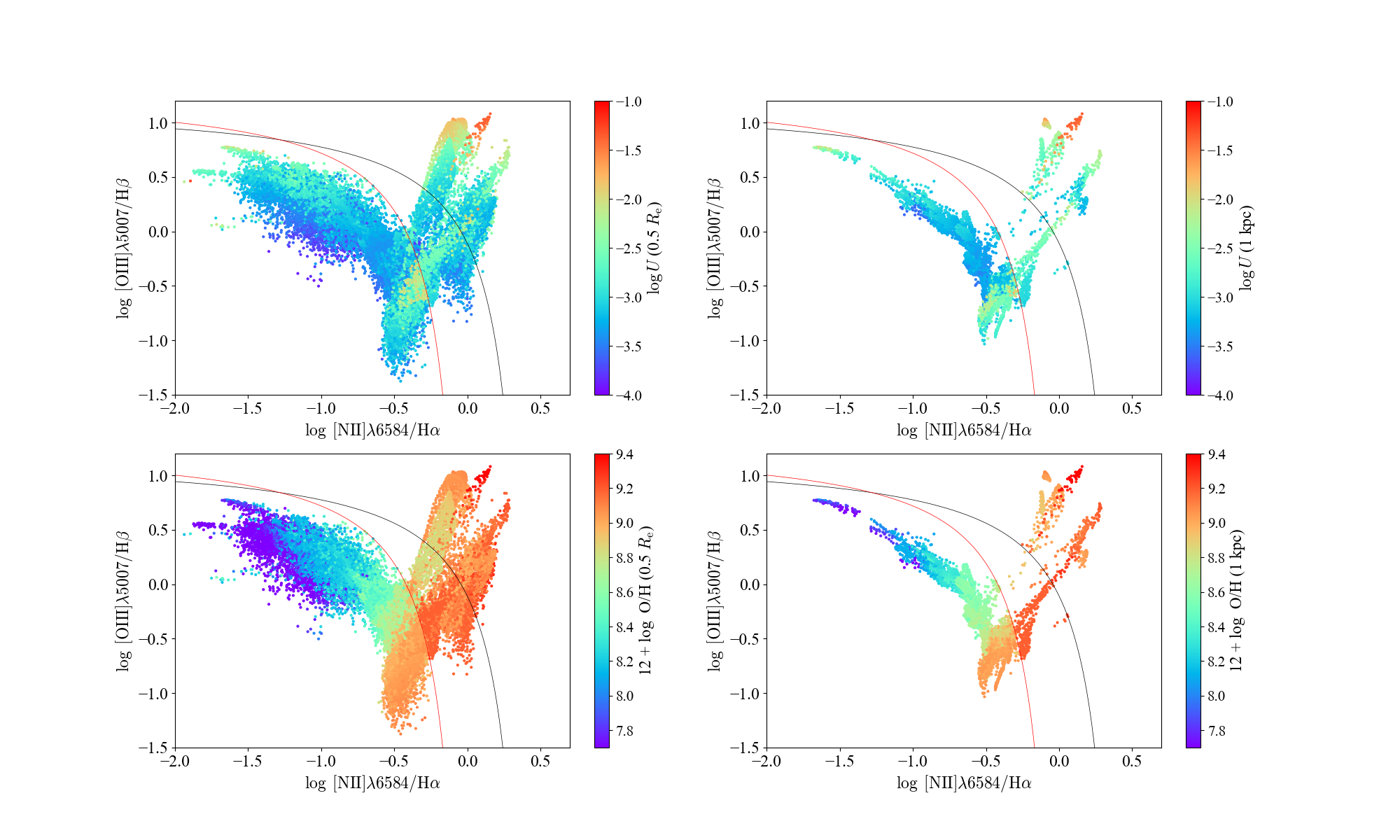}}%
    \caption{\nii-diagrams of the nuclear regions $r < 0.5 \ R_{\rm e}$ (left panels) and $r < 1$ kpc (right panels) centered on the SF and AGN host galaxies in the GASP sample. The points are color-coded according to the metallicity in the bottom panels, and according to the ionization parameter in the top panels. The red line is the \cite{kauffmann+2003} relation and the black line is the \cite{kewley+2001} relation.}
    \label{fig:bpt-diff-apert-gasp}
\end{figure}

\section{Ionization mechanism in the nuclear regions of AGN and SF galaxies from the GASP and MaNGA surveys} \label{sec:appendix-a0}

\begin{figure}[]
    \centering
    \makebox[\textwidth]{
    \includegraphics[scale=0.37]{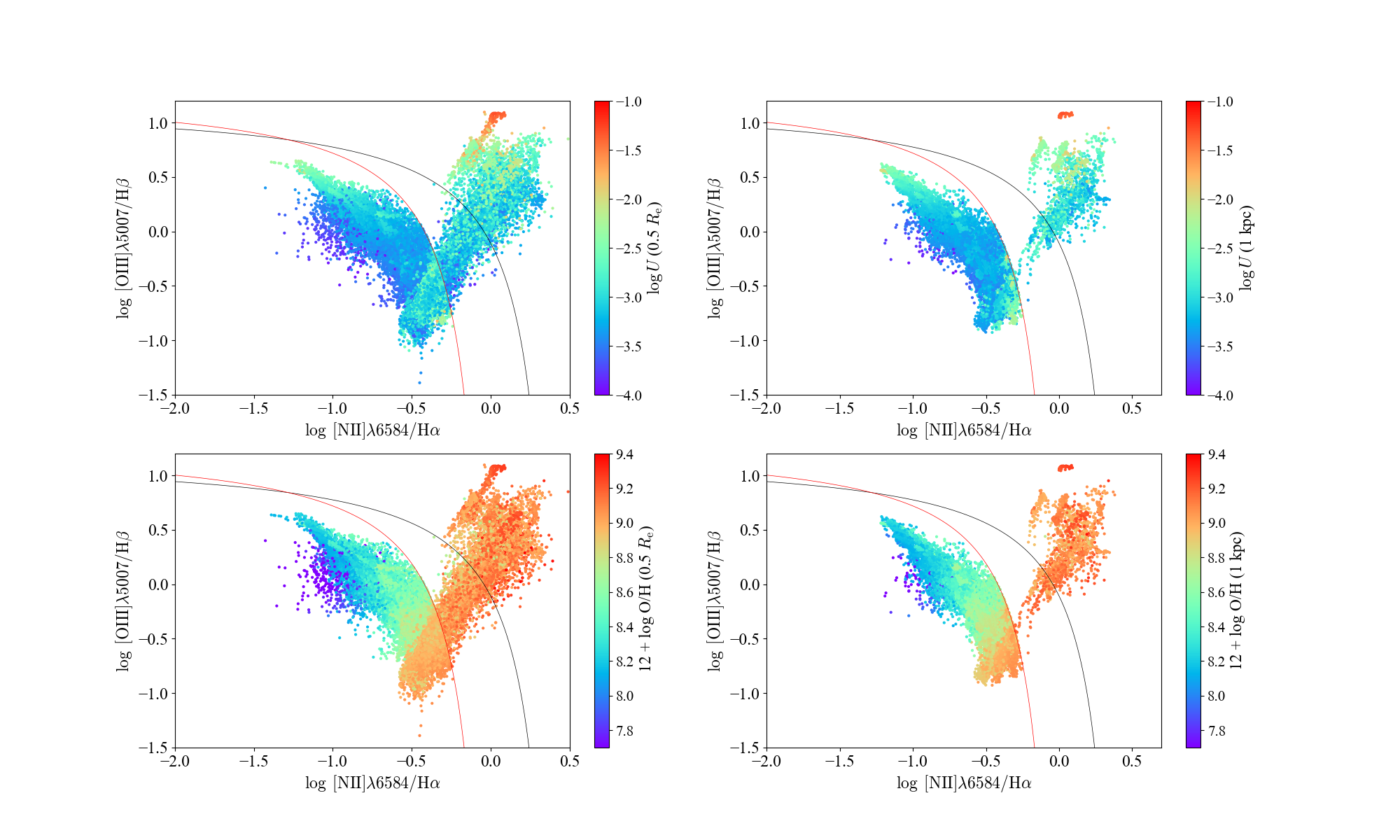}}%
    \caption{\nii-diagrams of the nuclear regions $r < 0.5 \  R_{\rm e}$ (left panels) and $r < 1$ kpc (right panels) centered on the SF and AGN host galaxies in the MaNGA sample.  The points are color-coded according to the metallicity in the bottom panels, and according to the ionization parameter in the top panels. The red line is the \cite{kauffmann+2003} relation and the black line is the \cite{kewley+2001} relation.}
    \label{fig:bpt-diff-apert-manga}
\end{figure}

Figures \ref{fig:bpt-diff-apert-gasp} and \ref{fig:bpt-diff-apert-manga} show the \nii-BPT diagrams of the emission line ratios within $r < 0.5 \ R_{\rm e}$ (left panels) and $r <$ 1 kpc (right panels) in the SF and AGN galaxies in the GASP and MaNGA samples, respectively. As mentioned in Section \ref{sec:results}, we compute the metallicity and ionization parameter inside the Composite spaxels of the AGN hosts galaxies, while we discard them in the SF galaxies as the ionization is presumably caused by mechanisms (such as e.g., shocks, DIG, diffuse X-ray emission) that are not taken into account in the present analysis. As a consequence, the density of points between the \cite{kewley+2001} and \cite{kauffmann+2003} relationships is relatively low with respect to the SF and AGN regions, especially in the plots showing the spaxels within the fixed aperture of $r <$ 1 kpc, which in general is smaller than the mass-scaled  regions defined by $r < 0.5 \ R_{\rm e}$.  
The color-coding shows that the ionization parameter and metallicity range between  $-4 \leq \log (U) \leq -1$ and $7.69 \leq 12 + \log \ {\rm (O/H)} \leq 9.4$, thus span a wide range of values even inside the nuclear regions of our galaxy samples.

\clearpage

\section{FMZR and Consistency between measurements from GASP and MaNGA}
\label{sec:appendix-c}
In Section \ref{sec:results-2} 
we have shown that the SF galaxies, from both the SF-FS and SF-RPS, are broadly located in the same region of the MZ diagram, even though showing a significant scatter along the y-axis. We discuss here how the scatter in the SF metallicities is mainly driven by the SFR in galaxies with $\log {\rm M}_* / {\rm M}_\odot < 10$ (before the plateau) in agreement with the FMZR \citep{mannucci+2010}.

\begin{figure}[ht]
    \centering
    \makebox[0.2\textwidth]{
    \includegraphics[scale=0.6]{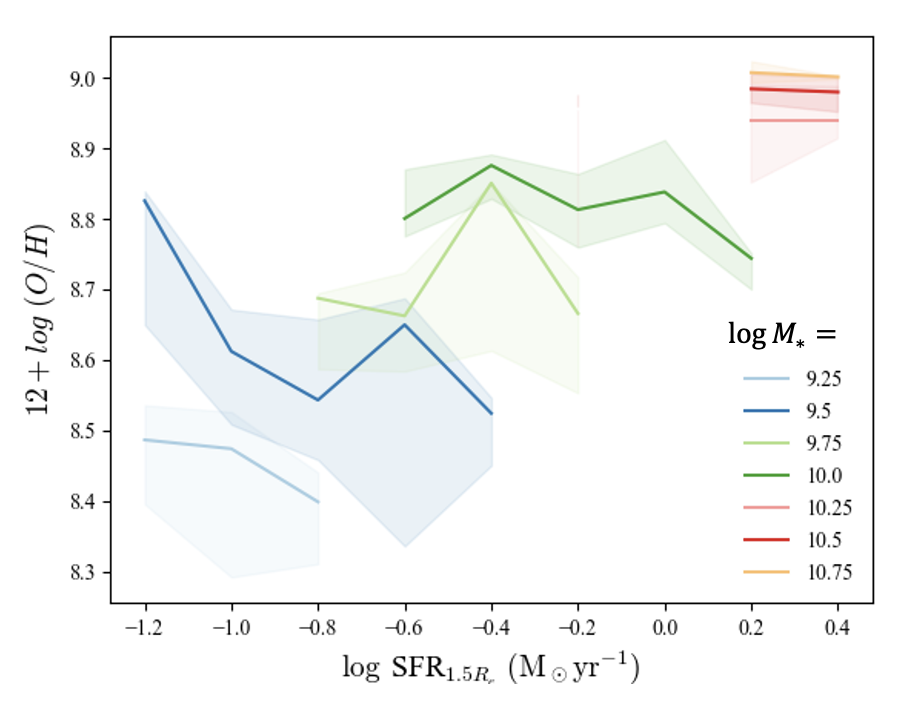}}%
    \caption{Median metallicity estimated inside the aperture of $r \sim$ 0.5 $R_{\rm e}$ as a function of the SFR (H$\alpha)_{1.5 R_{\rm e}}$ for the SF galaxies. We divide the galaxies into stellar mass bins of width 0.25 dex. The errors are the 25th and 68th percentile of the metallicity distribution inside each bin of mass. At $\log {\rm M}_* / {\rm M}_\odot \leq 9.5$, we observe an anti-correlation between the SFR(H$\alpha)_{1.5  R_{\rm e}}$ and 12 + log (O/H), with higher metallicities corresponding to lower SFR \citep[in agreement with][]{mannucci+2010}.}
    \label{fig:fmzr}
\end{figure}

Figure \ref{fig:fmzr} shows the median 12 + log (O/H) as a function of $\log$ SFR(H$\alpha)_{1.5 {\rm R_e}}$ of the SF galaxies, in different bins of stellar masses. We observe an anti-correlation between these two quantities in galaxies with  $\log ({\rm M}_*/{\rm M}_\odot) \leq 9.5$. For higher masses, instead,  
we observe again that metallicities saturate around the value 12 + log (O/H) $\sim$ 9.0 independently from the stellar mass and the SFR. 

In agreement with this, we observed in Figure \ref{fig:mzr} (Section \S \ref{sec:results-1}) that SF-RPS and SF-FS galaxies with $\log \ ({\rm M}_* / {\rm M}_\odot) < 10 $ observed by the GASP survey  have lower metallicities, but also higher SFR than the SF-FS galaxies from the MaNGA survey.



In particular, the SF-RPS  have a median SFR (H$\alpha)_{1.5Re}$ of 0.31$_{- 0.18}^{+0.32} \ {\rm M}_\odot \rm yr^{-1}$ and the SF-FS, considering the GASP galaxies only, have 0.38$_{-0.16}^{+ 0.15} \ {\rm M}_\odot \rm yr^{-1}$ at $\log \ ({\rm M}_* / {\rm M}_\odot) < 10 $, which are both higher than the median value of SFR (H$\alpha)_{1.5Re}$ = 0.14$_{-0.09}^{+0.20} \ {\rm M}_\odot \rm yr^{-1}$ for the MaNGA galaxies in the SF-FS, even if consistent within the errors. 
The median scatter of the GASP galaxies from the SF MZR is $\Delta$ (O/H) = 0.22 dex for the SF-RPS and $\Delta$ (O/H) = 0.14 dex for the SF-FS, in the mass range $\log {\rm M}_* / {\rm M}_\odot < 10$. 


We conclude that the MZR scatter in the SF sample is driven mainly by the SFR,
and that in particular the SF-FS galaxies from both the GASP and MaNGA surveys are located broadly in the same region of the diagram, confirming that there are no systematic effects linked to the surveys (such as different instruments, observations, data reduction) affecting the measurements of the metallicity or the stellar mass.

\end{document}